\documentclass[12pt]{article}
\usepackage{amsmath}
\usepackage{amsfonts} 
\usepackage{graphicx}
\usepackage{enumerate}
\usepackage{natbib}
\usepackage{url} 
\usepackage{hyperref}
\usepackage[english]{babel}
\usepackage{amsthm}
\usepackage{bbm}
\usepackage{soul}
\usepackage{paralist}
\usepackage{mathtools}
\usepackage{adjustbox}
\usepackage{booktabs}
\usepackage{xcolor}
\usepackage{amssymb}
\usepackage{multirow}  
\usepackage{array}     
\usepackage{physics}
\usepackage{enumitem}
\usepackage{subcaption}

\allowdisplaybreaks

\newtheorem{theorem}{Theorem}

\newtheorem{assumption}{Assumption}
\newtheorem{proposition}{Proposition}

\newtheorem{remark}{Remark}

\newcommand{\E}{\mathbb{E}}

\newcommand{\pr}{\mathrm{Pr}}

\DeclareMathOperator{\Cor}{Cor}

\usepackage{setspace}
\usepackage{bibspacing}
\newcommand{\blind}{0}

\begin{document}

	\def\spacingset#1{\renewcommand{\baselinestretch}%
		{#1}\small\normalsize} \spacingset{1}

	
	\if0\blind
	{
		\title{\bf 
			Principal stratification with recurrent events truncated by a terminal event: A nested Bayesian nonparametric approach
		}
		\author{Yuki Ohnishi  \\ 
			Department of Biostatistics, Yale School of Public Health, \\
			Michael O. Harhay \\
			Department of Biostatistics, Epidemiology and Informatics, \\ University of Pennsylvania, \\
			Guangyu Tong  \\ 
			Department of Biostatistics, Yale School of Public Health, \\
			and \\ 
			Fan Li\\
			Department of Biostatistics, Yale School of Public Health}

		\maketitle
	} \fi
	
	\if1\blind
	{
		\bigskip
		\bigskip
		\bigskip
		\begin{center}
			{\LARGE\bf 
				Principal stratification with recurrent events truncated by a terminal event: A nested Bayesian nonparametric approach
			}
		\end{center}
		\medskip
	} \fi
	
	\bigskip
	\begin{abstract}
		Recurrent events often serve as key endpoints in clinical studies but may be prematurely truncated by terminal events such as death, creating selection bias and complicating causal inference. To address this challenge, we develop a Bayesian nonparametric framework to address potential selection bias due to truncation by death within the continuous-time principal stratification framework. We introduce causal estimands for recurrent events in the presence of a terminal event and derive a partial identification result for the estimand under a dual-frailty framework, enabling transparent sensitivity analysis for non-identifiable parameters. We then propose a flexible Bayesian nonparametric prior, the enriched dependent Dirichlet process, specifically designed for joint modeling of recurrent and terminal events, addressing a limitation where standard Dirichlet process priors create random partitions dominated by recurrent events, yielding poor predictive performance for terminal events.  Simulations are carried out to show that our method has superior performance compared to existing methods. We apply the proposed new Bayesian nonparametric methods to infer the causal effect of a structured exercise program on rehospitalizations, which are subject to truncation by death.
	\end{abstract}

	\noindent%
	\vfill

	\newpage
	\spacingset{1.9} 
	
	\section{Introduction}
	\label{sec:intro}
	
	Clinical management of chronic heart failure often extends beyond guideline-directed medical therapy to include nonpharmacologic strategies such as exercise training and lifestyle modification. For stable patients with chronic systolic heart failure, exercise training is frequently recommended, but clinicians must weigh the feasibility, tolerance, and burden of a structured exercise prescription relative to potential benefits on clinically meaningful outcomes, in particular heart-failure hospitalizations and survival \citep{Heidenreich2022}. The HF-ACTION (Heart Failure: A Controlled Trial Investigating Outcomes of Exercise Training) trial was designed to address this question by testing whether adding supervised aerobic exercise training to usual care reduces all-cause mortality and hospitalization in patients with chronic systolic heart failure \citep{OConnor2009}. In practice, HF-ACTION speaks to a common outpatient dilemma: whether a formal exercise program can deliver measurable benefit on morbidity and mortality, without unacceptable safety concerns or poor adherence that limits effectiveness.
	
	Addressing this question requires moving beyond time-to-first hospitalization to quantifying how exercise affects the overall burden and timing of repeat hospital admissions. For chronic heart failure, patients commonly experience multiple rehospitalizations, so analyses restricted to a single ``first event'' can miss clinically meaningful differences in cumulative morbidity and disease burden. 
	The reduction in rehospitalization burden is also economically meaningful, because heart-failure hospitalizations are a major driver of overall spending \citep{Reed2010} and even modest decreases in recurrent admissions could translate into substantial per-patient and population-level cost savings.
	However, assessing the cumulative burden of rehospitalization can be complicated when the intervention also influences survival: longer survival increases exposure time for future hospitalizations, while earlier death both reduces time at risk and truncates the observation period for recurrent events like rehospitalization. 
	Further, when mortality differs by treatment, the observed recurrent-event process is observed under differential truncation, and comparisons can be distorted because the set of individuals who remain under observation later in follow-up is treatment-dependent. As a result, conventional regression approaches may conflate a treatment effect on the recurrent-event process with shifts in the composition of survivors, yielding estimates that reflect differential selection rather than a causal effect on the cumulative burden of rehospitalization. HF-ACTION exemplifies this broader challenge in evaluating chronic-disease management, where observed rehospitalizations are survivor-dependent while subject to differential truncation.

	While capturing the full history of hospitalizations provides a more complete picture of disease burden, analyzing such recurrent event processes introduces significant methodological challenges. The central issue here is how to define the causal effect of interest. In the presence of death, the phrase ``effect on rehospitalization'' is inherently ambiguous unless the target population and the counterfactual comparisons are clearly specified. A naive comparison of observed rehospitalization counts or rates between treatment arms fails to isolate the effect of exercise on the recurrent-event process, primarily because the treatment may simultaneously alter survival and, by extension, the opportunity to experience subsequent hospitalizations. Consequently, such comparisons conflate distinct phenomena: differences in survival time, differences in the frequency or timing of rehospitalization while alive, and shifts in the composition of the surviving cohort over time. To obtain a clinically interpretable causal effect, it is therefore necessary to explicitly define a common subpopulation for whom the rehospitalization burden is being compared and the specific counterfactual scenarios under which such a comparison remains meaningful.

	Principal stratification \citep{Frangakis2002} provides a general framework for causal inference that involves post-treatment intermediate variables and has gained prominence in addressing complexities caused by truncation due to death. 
	Within this framework, subpopulations, known as principal strata, are defined by intermediate potential outcomes under each treatment condition, enabling researchers to define unambiguous causal effects within these subgroups. 
	For example, the always-survivor subpopulation, consisting of patients who would survive until the non-terminal event occurs regardless of treatment assignment, was first introduced by \citet{Robins1986} and formally established by \citet{Zhang_Rubin_2003}. The associated identifiability has been previously studied when a non-mortality outcome is truncated by death, where death was typically considered as a simple, dichotomous event \citep[e.g.,][]{Long2013,Tchetgen_Tchetgen2014}.
	
	Extending the principal stratification framework to continuous time, a flourishing line of research has focused on evaluating the causal effect of treatment on a non-recurrent time-to-event outcome subject to semi-competing risks by time to death, or more generally, time to a terminal event \citep{Comment2025, Xu2022,Nevo2022}. By contrast, much less attention has been devoted to identifying causal effects on recurrent events subject to semi-competing risks by death, in which the recurrent event process may be informatively truncated by death. An exception is \citet{Lyu2023}, who proposed a causal estimand for the average number of recurrent events among always-survivors and developed a Bayesian parametric joint modeling approach to simultaneously represent the recurrent event and death processes. In their formulation, a single time index $t$ was used to define both the number of recurrent events and the always-survivor stratum. However, adopting a single time point $t$ to define causal estimands for recurrent event analysis with a terminal event can be less ideal because it considers the always-survivor population only at that specific time. This perspective makes the subpopulation under study vary by $t$, overlooking how treatment effects may accumulate or change over time for a fixed subpopulation and impeding comparisons of estimands across different time points.
	Finally, principal stratification typically requires careful parametric model specification that encompasses stringent regression assumptions and prior distributions. Therefore, the parametric modeling approach \citep[e.g.,][]{Nevo2022,Lyu2023,Comment2025} could be easily susceptible to model misspecification bias, especially under potentially complex data-generating processes.

	To address these challenges, we develop a Bayesian nonparametric framework for causal inference with recurrent events in the presence of a terminal event. Our novel contributions are severalfold.
	First, we propose new causal estimands and their identification results tailored for recurrent event analyses under semi-competing risks with death. Building on the double-index approach of \citet{Comment2025} to define always-survivor stratum, we derive partial identification results under a dual-frailty framework that cleanly separates within-arm (identifiable) dependence from cross-arm (non-identifiable) dependence. In particular, cross-world dependence is indexed by an interpretable sensitivity parameter, enabling a transparent sensitivity analysis of the cross-world dependence structure.
	Second, we develop a flexible Bayesian nonparametric (BNP) approach for inference. BNP methods have been increasingly used in recurrent event analysis to mitigate biases from model misspecification, a concern in many parametric approaches. In both causal and non-causal settings, previous work has employed Dirichlet process (DP) priors to jointly model recurrent and terminal event processes \citep[e.g.,][]{Paulon2020,Tian2024}. We discuss a potential pitfall in which a commonly used DP prior induces a random partition driven primarily by recurrent events, leading to inadequate learning about the terminal event component. To address this limitation, we introduce a nested DP structure and incorporate covariate dependence to relax the stringent exchangeability assumption. We refer to the resulting prior as the \emph{enriched dependent Dirichlet process} (EDDP), which also naturally accommodates the dual-frailty framework.
	We also develop a fully tractable and efficient Gibbs sampling algorithm for posterior computation and demonstrate through simulation studies that our method outperforms existing approaches in accuracy and robustness. 
	Finally, we analyze the HF-ACTION trial using the proposed methodology. Our analysis suggests that the formal exercise training reduces the expected number of rehospitalizations over long-term follow-up, with treatment effects becoming more pronounced over time. We show that the proposed estimand captures clinically relevant features of treatment benefit that conventional causal estimands do not fully reflect in the HF-ACTION setting. In addition, by varying the sensitivity parameter within a Bayesian g-computation scheme, we transparently assess how causal conclusions change under alternative cross-world dependence structures. The main conclusions remain stable across small to moderately large values of the sensitivity parameter. Only under an extreme dependence assumption, where the proposed dual-frailty formulation effectively collapses to a single-frailty approach, does the evidence become attenuated, highlighting the importance of modeling distinct cross-world frailties in this setting.

	\subsection{HF-ACTION (Heart Failure: A Controlled Trial Investigating Outcomes of Exercise Training)}
	\label{sec:motivating_example}
	
	HF-ACTION was a multicenter randomized clinical trial designed to evaluate whether adding a structured aerobic exercise program to guideline-based medical therapy reduces clinical events in patients with chronic systolic heart failure \citep{OConnor2009}. Medically stable outpatients with left ventricular ejection fraction (LVEF) $\leq 35\%$ and New York Heart Association (NYHA) class II--IV symptoms despite optimal therapy were randomized 1:1 to usual care alone or usual care plus guided aerobic exercise training. 
	The exercise intervention consisted of a structured supervised phase, with a goal of 36 sessions over approximately 3 months, followed by continued home-based training with ongoing adherence monitoring and encouragement during follow-up; the protocol does not indicate that exercise training stopped after a first rehospitalization. The usual-care arm did not receive a formal exercise prescription. In the original HF-ACTION trial, the primary endpoint was the composite of all-cause mortality or all-cause hospitalization, analyzed as a time-to-first-event outcome, with prespecified secondary endpoints of all-cause mortality, cardiovascular mortality or cardiovascular hospitalization, and cardiovascular mortality or heart-failure hospitalization. Thus, our focus on rehospitalization burden extends the original trial perspective beyond first-event analysis.
	In this work, the outcome of interest is the sequence of hospitalization events recorded during 4 years of follow-up, and death is treated as an absorbing terminal event that truncates observation of subsequent hospitalizations. This structure motivates estimands that go beyond time-to-first-event: for chronic heart failure, the cumulative burden and timing of repeated hospitalizations are clinically meaningful, but are entangled with survival because longer survival both increases time at risk for hospitalization and reduces truncation of the recurrent process by death.
	
	Table~\ref{tab:hfaction_table1} summarizes baseline covariates, events during follow-up, and distribution of rehospitalization by treatment arms in the analytic dataset. As expected in a randomized trial, baseline characteristics are well balanced across arms, including key prognostic factors such as left ventricular ejection fraction (LVEF), New York Heart Association (NYHA) class, and cardiopulmonary exercise (CPX) test duration. Mean follow-up is 928 days averaged across both arms, reflecting common trial protocol and administrative censoring. Hospitalizations are common and recurrent: more than 50\% of participants experience at least one hospitalization, and the distribution of rehospitalization exhibits both excess zeros and a right tail (maximum 12 events). The gap-time summaries in Table~\ref{tab:hfaction_table1} show substantial heterogeneity in the spacing of subsequent events, but the gap times decrease sharply for later recurrences. As a result, inference about late-event dynamics will be driven by a small subset with frequent hospitalizations.
	
	Table~\ref{tab:hfaction_table2} focuses on the clinically important subset of participants with at least one hospitalization. Two patterns are especially relevant for our analyses. First, within this subset, death occurs in a substantial portion of participants: more than 20\% die during follow-up, suggesting that the rehospitalization process may beinformatively truncated by mortality among higher-risk patients. Second, among decedents the last observed hospitalization often occurs closer to death than the corresponding time from the last hospitalization to non-death censoring among survivors, reinforcing the interdependence of rehospitalization burden and mortality.
	
	These empirical features highlight why causal analysis of rehospitalizations in HF-ACTION cannot rely solely on naive comparisons of observed hospitalization counts or rates between treatment arms, even under randomization. Because exercise training may affect both hospitalization and survival, and death censors future hospitalizations, the observed recurrent-event process mixes (i) the causal effect of treatment on hospitalization dynamics while alive and (ii) the causal effect of treatment on survival (and hence on time at risk and truncation). To separate these mechanisms, our methodology adopts a principal stratification perspective tailored to recurrent outcomes: we target treatment effects within a subpopulation whose recurrent histories up to a chosen horizon $r$ would not be truncated by death under either treatment assignment. For inference, we couple this estimand with a flexible Bayesian framework that jointly models hospitalization gap times and the terminal event while accommodating strong heterogeneity in event processes. Finally, because cross-arm dependence between potential outcomes is fundamentally unidentifiable, we employ a joint-frailty sensitivity formulation that separates identifiable within-arm dependence (between hospitalization and death) from non-identifiable cross-arm dependence, enabling transparent assessment of how causal conclusions shift under alternative cross-world dependence structures.

	\begin{table}[!htbp]
		\centering
		\small
		\caption{Baseline covariates, follow-up, and hospitalization (recurrent-event) summaries by treatment arm in the HF-ACTION trial. Binary covariates and count-based summaries are shown as $n$ (\%), continuous summaries as mean (SD) or median [Q1,Q3]. Hospitalization and follow-up times are measured in days from randomization. Gap-time summaries condition on having at least $j$ events and report the gap ending in the $j$th event.}
		\label{tab:hfaction_table1}
		\begin{adjustbox}{width=14cm}
			\begin{tabular}{lrr}
				\toprule
				Characteristic & Usual care (n=1070) & Exercise training (n=1060) \\
				\midrule
				\multicolumn{3}{l}{\textbf{Baseline covariates}} \\
				Age (years) & 58.51 (12.97) & 58.60 (12.38)\\
				Female & 281 (26.3\%) & 318 (30.0\%) \\
				BMI (kg/m$^2$) & 31.04 (7.08) & 30.95 (7.16) \\
				NYHA class III/IV & 379 (35.4\%) & 400 (37.7\%) \\
				LVEF (\%) & 25.23 (7.36) & 25.19 (7.60) \\
				Ischemic etiology & 546 (51.0\%) & 550 (51.9\%) \\
				Angina & 250 (23.4\%) & 277 (26.1\%) \\
				Prior MI & 455 (42.5\%) & 444 (41.9\%) \\
				Stroke history & 113 (10.6\%) & 105 (9.9\%) \\
				History of diabetes & 335 (31.3\%) & 343 (32.4\%) \\
				History of depression & 237 (22.1\%) & 215 (20.3\%) \\
				Systolic BP (mm Hg) & 113.98 (18.57) & 113.57 (17.92) \\
				Diastolic BP (mm Hg) & 70.15 (11.26) & 70.27 (11.32) \\
				Baseline heart rate (bpm) & 70.82 (11.66) & 70.95 (11.25) \\
				CPX test duration & 9.88 (4.03) & 9.74 (3.80) \\
				\addlinespace
				\multicolumn{3}{l}{\textbf{Follow-up and event summaries}} \\
				Follow-up time (days) & 923.9 (400.8) & 931.3 (388.3) \\
				Death observed & 183 (17.1\%) & 168 (15.8\%) \\
				Any hospitalization & 594 (55.5\%) & 561 (52.9\%)\\
				Hospitalizations per participant & 0.922 (1.198) & 0.868 (1.108)\\
				Total hospitalizations & 987 & 920 \\
				Hospitalization rate (per 100 person-years) & 36.46 & 34.04 \\
				Days to first hospitalization among those with $\geq 1$ event & 293 [128, 573] & 275 [120, 544] \\
				\addlinespace
				\multicolumn{3}{l}{\textbf{Distribution of recurrent events per participant}} \\
				0 hospitalizations & 476 (44.49\%) & 499 (47.08\%) \\
				1 hospitalization & 360 (33.64\%) & 341 (32.17\%) \\
				2 hospitalizations & 143 (13.36\%) & 132 (12.45\%) \\
				3 hospitalizations & 64 (5.98\%) & 62 (5.85\%) \\
				4 hospitalizations & 12 (1.12\%) & 11 (1.04\%) \\
				5 hospitalizations & 7 (0.65\%) & 9 (0.85\%) \\
				6 hospitalizations & 1 (0.09\%) & 3 (0.28\%) \\
				7+ hospitalizations & 7 (0.65\%) & 3 (0.28\%) \\
				\addlinespace
				\multicolumn{3}{l}{\textbf{Gap times between recurrent events (days)}} \\
				Gap between 1st and 2nd events & 132 [39, 333] & 107 [32, 363] \\
				Gap between 2nd and 3rd events & 138 [58, 292] & 106 [35, 244] \\
				Gap between 3rd and 4th events & 127 [49, 195] & 86 [49, 128] \\
				Gap between 4th and 5th events & 99 [64, 284] & 67 [46, 200] \\
				Gap to 6th+ event (from previous event) & 78 [38, 100] & 60 [26, 94] \\
				\addlinespace
				\bottomrule
			\end{tabular}
		\end{adjustbox}
	\end{table}
	
	\begin{table}[!htbp]
		\centering
		\small
		\caption{Summary among participants with at least one observed hospitalization, stratified by treatment arm in HF-ACTION. Cells report mean (SD) or median [Q1,Q3].}
		\label{tab:hfaction_table2}
		\begin{adjustbox}{width=14cm}
			\begin{tabular}{lrr}
				\toprule
				Characteristic & Usual care (n=594) & Exercise training (n=561) \\
				\midrule
				\multicolumn{3}{l}{\textbf{Sample sizes and event burden (among those with $\geq 1$ recurrent event)}} \\
				Deaths in this subset & 133 (22.4\%) & 122 (21.7\%) \\
				Censoring (non-death) in this subset & 461 (77.6\%) & 439 (78.3\%) \\
				Hospitalizations per participant & 1.66 (1.17) & 1.64 (1.03) \\
				\addlinespace
				\multicolumn{3}{l}{\textbf{Time from last hospitalization to end-of-follow-up (days)}} \\
				Last hospitalization to death: median [Q1,Q3] & 124 [25, 332] & 165 [16, 380]\\
				Last hospitalization to death: mean (SD) & 221.6 (255.7) & 256.6 (289.2) \\
				Last hospitalization to censoring: median [Q1,Q3] & 436 [212, 799] & 494 [232, 825] \\
				Last hospitalization to censoring: mean (SD) & 523.6 (373.7) & 544.9 (369.7)  \\
				\addlinespace
				\bottomrule
			\end{tabular}
		\end{adjustbox}
	\end{table}
	\section{Notation, Setup, and Estimands}
	\label{sec:setup}
	\subsection{Potential outcomes and observed data structure}
	\label{sec:notation}
	We consider a study involving $n$ units, each assigned to one of two treatment arms at the beginning of the study:
	$Z_i = 1$ for the treatment group and $Z_i = 0$ for the control group.
	Let $\mathbf{X}_i \in \mathcal{X}$ denote a vector of baseline covariates for unit $i$.
	For unit $i$, let $D_i^z$ denote the potential time to the terminal event under treatment $z \in \{0,1\}$.
	Throughout the paper, the terminal event is death; more generally, $D_i^z$ may represent any absorbing terminal event that truncates recurrent follow-up.
	In the motivating HF-ACTION analysis, $D_i^z$ corresponds to all-cause mortality, and the recurrent outcome consists of hospitalizations so that death serves as the truncating terminal event.
	Let $C_i^z$ denote the potential right-censoring time under treatment $z$ due to loss to follow-up or administrative end of study (i.e., any censoring mechanism other than death).
	The observed follow-up time is
	$\mathcal{T}_i = \min(D_i^{Z_i}, C_i^{Z_i})$.
	Define the indicators $\delta_i^D = \mathbbm{1}\{\mathcal{T}_i = D_i^{Z_i}\}$,
	$\delta_i^C = \mathbbm{1}\{\mathcal{T}_i = C_i^{Z_i}\}$
	so that $\delta_i^D=1$ indicates that death is observed and $\delta_i^C=1$ indicates right-censoring for reasons other than death.

	We assume that the potential cumulative number of recurrent events for subject $i$ under treatment $z$, $N_i^z(\cdot)$, follows a point process. 
	The potential point process $N_i^z(t)$ is well-defined only for $0 \leq t < D_i^z$ and undefined for $t\geq D_i^z$ due to truncation; that is, we use  $N_i^z(t)=*$ for the undefined case following the convention considered in \citet{Zhang_Rubin_2003}. Let $N_i=N_i^{Z_i}(\mathcal{T}_i-)$ denote the observed number of recurrent events until the last follow-up.
	Let $T_{ij}^z$ represent the time of the $j$-th recurrent event for subject $i$ under treatment $z$ for any $j \in \mathbb{N}$. 
	Since the recurrent event is truncated for those who experience the terminal event, we also define $T_{ij}^z=*$ for all $j$ with $T_{ij}^z>D_{i}^z$. As a special case, we write $T^z_{i0}=0$.
	The observed event times are $T_{ij} = T_{ij}^{Z_i}$, which are observed only for $j$ with $T_{ij} \leq \mathcal{T}_i$. 
	The observed event times are $\{T_{ij}\}_{j=1}^{N_i}$, and the gap time between two successive events is defined as $W_{i1}^z = T_{i1}^z$ and $W_{ij}^z = T_{ij}^z - T_{i(j-1)}^z,$ for $j \geq 2$. For the observed data, the gap times are $W_{ij} = T_{ij} - T_{i(j-1)}$. 
	Since the time to event distribution is often right-skewed, we work on the log-time scale later in Section~\ref{sec:BNPmodel}.  Define the log-scale terminal time $U_i^z=\log(D_i^z)$
	and the log-scale gap times $Y_{ij}^z=\log(W_{ij}^z)$ $j\in\mathbb N^{+}$.
	Because $U_i^z$ and $Y_{ij}^z$ are one-to-one transformations of $D_i^z$ and $W_{ij}^z$, modeling $(U_i^z, Y_{ij}^z)$ is equivalent to modeling $(D_i^z, W_{ij}^z)$, and follows the typical convention in the survival analysis literature.
	This explicitly connects the model in Section~\ref{sec:BNPmodel} to the causal estimands defined in Section \ref{sec:estimands}.
	Finally, the observed data for each unit consist of the tuple $O_i=\{\mathcal{T}_i,\delta_i^D,\delta_i^C,N_i,\{T_{ij}\}_{j=1}^{N_i},Z_i,\mathbf{X}_i \}$.
	We next state assumptions under which the causal estimands are identified from the observed data
	(up to the standard cross-world dependence addressed via sensitivity analysis in Section~\ref{sec:bayesian_inference}).
	
	\begin{assumption}[Stable Unit Treatment Value Assumption (SUTVA)]
		\label{asmp:consistency}
		There is no interference between units and no hidden versions of treatment.
		For each unit $i$, if $Z_i=z$ then $D_i = D_i^z$, $C_i=C_i^z$, and $N_i(t)=N_i^z(t)$ for all $t \ge 0$.
		Equivalently, for the transformed variables, $U_i=\log(D_i)=U_i^z$ and $Y_{ij}=\log(W_{ij})=Y_{ij}^z$ under $Z_i=z$.
	\end{assumption}
	\begin{assumption}[Treatment ignorability]
		\label{asmp:ignorability} 
		For any follow-up time $t$, we have $Z_i \perp \{N_i^0(t),  D_i^0, N_i^1(t) , D_i^1\}   \mid \mathbf{X}_i.$
	\end{assumption}
	\begin{assumption}[Covariate-dependent censoring]
		\label{asmp:indep_censoring}
		For any $t$, we have $\{ C_i^0, C_i^1 \}  \perp \{N_i^0(t), D_i^0, N_i^1(t), D_i^1\} \mid \mathbf{X}_i$.
	\end{assumption}
	Assumptions~\ref{asmp:consistency} and~\ref{asmp:ignorability} are standard in randomized trials and observational studies.
	Assumption~\ref{asmp:consistency} encodes SUTVA (no interference and well-defined treatments), and
	Assumption~\ref{asmp:ignorability} formulates the treatment ignorability, which is guaranteed to hold under randomization in HF-ACTION but more generally applies to observational studies with unconfounded treatment assignment.
	Due to the nature of the intervention and randomization, both assumptions hold in our motivating HF-ACTION trial.
	Assumption~\ref{asmp:indep_censoring} states that potential censoring times are conditionally independent of the potential event outcomes.
	Although empirically unverifiable due to the cross-world counterfactuals, this assumption is reasonable in HF-ACTION because the majority of right-censored events were due to the pre-specified end of follow-up.

	\subsection{Principal stratum causal estimands in continuous time}
	\label{sec:estimands}

	Principal stratification \citep{Frangakis2002} is a general framework for addressing intermediate outcomes and is particularly attractive for applications with death truncation. This framework focuses on the always-survivor subgroup defined by the combination of observed and missing potential outcomes for the terminal event. Suppose the time to the first event is of interest, an existing estimand is the survivor average causal effect (SACE) for a fixed time point $t$: 
	$$\pr(T_{i1}^1 < t \mid t <D_{i}^0, t <D_{i}^1 ) - \pr(T_{i1}^0 < t \mid t <D_{i}^0, t <D_{i}^1 ).$$  This quantity allows us to circumvent the potential bias by focusing on the subpopulation of individuals who would always survive regardless of the assigned treatment, i.e., both the first and second probability statements are conditioned on the same cohort of individuals. For recurrent event analysis, \citet{Lyu2023} adopted this definition of the always-survivor and defined the SACE for recurrent events with a terminal event. In particular, they considered the causal contrast of the number of recurrent events defined as 
	$$\frac{\E[N_{i}^{1}(t) \mid t <D_{i}^0, t <D_{i}^1]}{\E[ N_{i}^{0}(t) \mid t <D_{i}^0, t <D_{i}^1]}.$$

	\citet{Comment2025} argued that estimating the SACE at a single time point $ t $ may present certain challenges. This is because it describes causal effects only for the always-survivor subpopulation at that specific moment, which they call a \emph{snapshot effect}. This static approach fails to account for the sensitivity to the choice of $ t $ because the focused subpopulation changes and diminishes over time, and thus overlooks how treatment effects accumulate or vary over time for a fixed subpopulation. Consequently, these snapshot estimands cannot capture the full temporal nature of treatment effects and fail to provide meaningful insights about the causal effects.
	\citet{Comment2025} addressed this issue by focusing on double-indexed estimands, where the principal strata are defined by the time index $ r $, independent of the choice of $ t $ that indexes the time to the events of interest. This approach introduces the time-varying SACE (TV-SACE), where the population of always-survivors depends only on the chosen time $ r $, that is, 
	$$\pr(T_{i1}^1 < t \mid r <D_{i}^0, r <D_{i}^1 ) - \pr(T_{i1}^0 < t \mid r <D_{i}^0, r <D_{i}^1 ).$$ Their estimands focus on the first time-to-event outcome rather than recurrent events.

	We adopt the double-index approach for the recurrent event analysis and define the subpopulation of always-survivors at time $r$ as: $$\mathcal{AS}(r) = \{i: r <D_{i}^0, r <D_{i}^1 \} ~~\text{for}~~ t < r. $$
	Our interest lies in evaluating the causal effect on the recurrent event process for those in $\mathcal{AS}(r)$, which is relevant when $N_i^z(\cdot)$ and $T_{ij}^z$ can be undefined due to truncation. 
	We define the \emph{survivor-average number of recurrence} (SANR) estimands as
	\begin{equation}
		\label{eq:estimand_1}
		\text{SANR}(t; r)  = g\qty{\mu^1\qty(t;r), \mu^0\qty(t;r)},
	\end{equation}
	where $\mu^z\qty(t;r) = \E\qty[N_i^z(t) \mid \mathcal{AS}(r)]$ for $t<r$. The function $g(\cdot,\cdot)$ determines the scale of the effect measure. For example, $g(x,y)=x-y$ and $g(x,y)=x/y$ correspond to causal mean difference and causal rate ratio, respectively. 
	This class of estimands represents the causal contrast in the number of event occurrences before time $t$ for the subset of patients who would survive until time $r$ under both treatment arms. This estimand generalizes those considered in \citet{Lyu2023}, to which it reduces when $r=t$ and $g(x,y)=x/y$.
	Table \ref{tab:summary_ps_literature} summarizes how our definitions of the estimands and the always-survivor stratum differ from those in the existing literature, thereby positioning the contribution of this paper relative to prior work on principal stratification.
	
	\begin{table}
		\centering
		\caption{Summary of literature on principal stratification in continuous time. We summarize the following features: (i) type of outcome, (ii) scale of estimand, (iii) definition of always-survivor stratum, (iv) cross-world assumption (v) whether cross-world and within-world dependence is distinguished, (vi) identification strategy (point identification vs partial identification with sensitivity analysis (SA)), and (vii) inference model. }
		\begin{adjustbox}{width=14.2cm}
			\begin{tabular}{l rrrrr}
				\toprule
				& \citet{Nevo2022} & \citet{Xu2022} & \citet{Lyu2023} & \citet{Comment2025} & This article  \\
				\midrule
				\emph{Outcome type           } & Time to event             & Time to event   & Recurrent event & Time to event & Recurrent event  \\
				\emph{Estimand scale         } & Difference                & Ratio           & Ratio           & Difference    & General scale  \\
				\emph{Always-survivor stratum} & Population stratification & Single index    & Single index    & Double index  & Double index \\
				\emph{Cross-world assumption } & Frailty                   & Copula          & Frailty         & Frailty       & Frailty \\
				\emph{Cross- and within-world} & Distinguished             & Distinguished   & Conflated       & Conflated     & Distinguished \\
				\emph{Identification         } & Partial with SA           & Partial with SA & Point           & Point         & Partial with SA \\
				\emph{Model                  } & Parametric                & Nonparameteric  & Parametric      & Parametric    & Nonparameteric \\
				\bottomrule
			\end{tabular}
		\end{adjustbox}
		\label{tab:summary_ps_literature}
	\end{table}

	\section{Bayesian Nonparametric Causal Inference}
	\label{sec:bayesian_inference}
	\subsection{Overview} 
	Section~\ref{sec:setup} defined the potential terminal time $D_i^z$ and the recurrent-event schedule
	$\{ T_{ij}^z\}_{j\ge 1}$ (equivalently $ N_i^z(\cdot)$) under each arm $z\in\{0,1\}$.
	We work on the log-time scale in our Bayesian model: $U_i^z=\log(D_i^z)$ and $Y_{ij}^z=\log( W_{ij}^z)$
	with $ W_{ij}^z= T_{ij}^z- T_{i(j-1)}^z$.
	If unit $i$ is assigned to arm $Z_i=z$, then each observed gap time $W_{ij}=T_{ij}-T_{i(j-1)}$ corresponds to the $j$th gap under arm $z$,
	so that by Assumption~\ref{asmp:consistency} we observe
	$Y_{ij}=\log(W_{ij})=Y_{ij}^z=\log( W_{ij}^z)$ for $j=1,\ldots,N_i$.
	The next  gap under the assigned arm, $Y_{i,N_i+1}=Y_{i,N_i+1}^{Z_i}$, is right-censored at
	$L_{i,N_i+1}=\log(\mathcal{T}_i-T_{iN_i})$ because no additional recurrence is observed in $(T_{iN_i},\mathcal{T}_i]$;
	this censoring is induced either by the terminal event ($\mathcal{T}_i=D_i^{Z_i}$) or by other censoring mechanisms ($\mathcal{T}_i=C_i^{Z_i}$).
	Likewise, the terminal log-time under the assigned arm, $U_i=U_i^{Z_i}=\log(D_i^{Z_i})$, is observed when $\delta_i^C=0$ (death observed) and is right-censored at $\log(\mathcal{T}_i)$ when $\delta_i^C=1$.
	Therefore, the observed-data likelihood (defined later in Section~\ref{sec:BNPmodel}) is exactly the likelihood for the  variables $\{U_i^{Z_i},Y_{i1}^{Z_i},Y_{i2}^{Z_i},\ldots\}$ under this right-censoring structure.
	Because $(U_i^z,\{Y_{ij}^z\}_{j\ge 1})$ is a one-to-one transformation of $(D_i^z,\{ T_{ij}^z\}_{j\ge 1})$,
	a model for $(U_i^z,\{Y_{ij}^z\}_{j\ge 1})$ induces a model for $(D_i^z, N_i^z(\cdot))$ and thus for the causal estimands.

	Let $\mathbf{Y}_i^z=(Y_{i1}^z,Y_{i2}^z,\ldots)$ denote the infinite log-gap sequence under arm $z$.
	Bayesian inference targets the joint distribution of all variables
	$(\mathbf{Y}^0,\mathbf{U}^0,\mathbf{Y}^1,\mathbf{U}^1,\mathbf{C}^0,\mathbf{C}^1,\mathbf{Z},\mathbf{X})$.
	We assume \emph{a priori} independence between parameters governing $(\mathbf{C}^0,\mathbf{C}^1,\mathbf{Z},\mathbf{X})$
	and those governing $(\mathbf{Y}^0,\mathbf{U}^0,\mathbf{Y}^1,\mathbf{U}^1)$.
	Under Assumption~\ref{asmp:indep_censoring} and this prior independence, the censoring mechanism is ignorable for inference
	about the recurrent/terminal potential outcomes (e.g., \citealp{Rubin1978}), so we focus on
	$(\mathbf{Y}^0,\mathbf{U}^0,\mathbf{Y}^1,\mathbf{U}^1,\mathbf{Z},\mathbf{X})$.
	
	Assuming unit exchangeability, de Finetti’s theorem implies the existence of global parameters
	$\boldsymbol{\theta}^*$ and $\boldsymbol{\varphi}^*$ such that
	\begin{equation}
		\label{eq:joint_dist}
		\begin{split}
			&p\!\left(\mathbf{Y}^0,\mathbf{U}^0,\mathbf{Y}^1,\mathbf{U}^1,\mathbf{Z},\mathbf{X}\right) \\
			&\qquad=
			\int p(\boldsymbol{\theta}^*)\,p(\boldsymbol{\varphi}^*)
			\prod_{i=1}^{n}
			p\!\left(\mathbf{Y}_i^0,U_i^0,\mathbf{Y}_i^1,U_i^1 \mid \mathbf{X}_i,\boldsymbol{\theta}^*\right)
			\,p\!\left(Z_i,\mathbf{X}_i\mid \boldsymbol{\varphi}^*\right)
			\,d\boldsymbol{\theta}^*\,d\boldsymbol{\varphi}^*.
		\end{split}
	\end{equation}
	where the global parameter $\boldsymbol{\theta}^{*}$ has prior $p(\boldsymbol{\theta}^{*})$.
	Under the ignorable treatment assignment (Assumption \ref{asmp:ignorability}), the assignment model does not affect the inference of the quantities of scientific interest $\boldsymbol{\theta}^{*}$ either. In what follows, we will conduct a likelihood analysis for $\boldsymbol{\theta}^{*}$, assuming that the value of $\boldsymbol{\theta}^{*}$ which governed the distribution of observable data has been drawn from a prior distribution with compact support. For identification, we further assume that a \emph{frailty} $\boldsymbol{\gamma}$ is included as unknown parameters in an expanded parameter set $\boldsymbol{\theta}^{*}=(\boldsymbol{\theta}, \boldsymbol{\gamma})$ for $\boldsymbol{\gamma}=(\boldsymbol{\gamma}_1, \ldots, \boldsymbol{\gamma}_n)^\top$, as formalized in the following section.

	\subsection{Identification via frailty and sensitivity analysis}

	The causal estimands depend on the joint distribution of the potential outcomes across arms,
	which is not identified from observed data alone because $(\mathbf{Y}_i^0,U_i^0)$ and $(\mathbf{Y}_i^1,U_i^1)$ are never jointly observed. To resolve this identification challenge, we introduce a bivariate frailty $\boldsymbol{\gamma}_i=(\gamma_i^0,\gamma_i^1)$ to transparently parameterize cross-world dependence.
	
	\begin{assumption}[Cross-world dependence via bivariate frailty]
		\label{asmp:indep_given_fraity}
		For each unit $i$, there exists a bivariate frailty $\boldsymbol{\gamma}_i=(\gamma_i^0,\gamma_i^1)$ such that 
		\begin{align*}
			p\!\left(\mathbf{Y}_i^0,U_i^0,\mathbf{Y}_i^1,U_i^1 \mid \mathbf{X}_i,\boldsymbol{\gamma}_i\right)
			&= p\!\left(\mathbf{Y}_i^0,U_i^0 \mid \mathbf{X}_i,\gamma_i^0\right)\; p\!\left(\mathbf{Y}_i^1,U_i^1 \mid \mathbf{X}_i,\gamma_i^1\right)   \text{ and }\\
			p\!\left(\mathbf{Y}_i^z,U_i^z \mid \mathbf{X}_i,\gamma_i^z\right)
			&=p\!\left(\mathbf{Y}_i^z \mid \mathbf{X}_i,\gamma_i^z\right)p\!\left(U_i^z \mid \mathbf{X}_i,\gamma_i^z\right) \text{ for } z=0,1.
		\end{align*} 
	\end{assumption}
	\noindent
	Although the frailty device has been previously adopted in the semi-competing risks literature \citep[e.g.,][]{Stensrud2017,Nevo2022,Lyu2023,Comment2025}, our approach differs from the existing literature in several important respects.
	Table \ref{tab:summary_ps_literature} summarizes the key distinctions between our frailty-based approach and prior work.
	
	Under Assumption \ref{asmp:indep_given_fraity}, the individual frailty vector $\boldsymbol{\gamma}_i=(\gamma_i^0,\gamma_i^1)$ governs two distinct forms of dependence: (i) \emph{cross-world} dependence between treatment arms, arising from the association between $\gamma_i^0$ and $\gamma_i^1$; and (ii) \emph{within-world} dependence between the terminal event $U_i^z$ and the recurrent event process $\mathbf{Y}_i^z$ under treatment $z$. Component (i) is unidentifiable because $(\mathbf{Y}_i^0,U_i^0)$ and $(\mathbf{Y}_i^1,U_i^1)$ are never jointly observable, whereas Component (ii), which captures within-world dependence between recurrent events and the terminal event, has been recognized as an important source of variation in noncausal, associative analyses, where cross-world dependence is not considered \citep[e.g.,][]{Paulon2020,Tian2024}.

	A common simplification sets $\gamma_i^0=\gamma_i^1\equiv\gamma_i$ \citep[e.g.,][]{Lyu2023,Comment2025}, thereby assuming cross-world independence in Component (i) and yielding point identification of the causal contrasts. In causal analyses, however, this restriction also imposes a common within-world dependence structure between recurrent events and the terminal event across treatment worlds, an assumption that is unlikely to hold unless all effect modifiers are correctly measured and modeled. Therefore, this approach conflates two distinct dependence structures, one inherently non-identifiable (Component (i)) and the other identifiable (Component (ii)), thereby imposing a strong and opaque assumption in causal settings.
	
	Our approach retains separate frailties $\gamma_i^0$ and $\gamma_i^1$ and introduces a sensitivity parameter $\rho=\Cor(\gamma_i^0,\gamma_i^1)$ that captures their cross-world association, similar to \citet{Nevo2022} who addressed the non-recurrent event setting.  
	This formulation cleanly separates identifiable and non-identifiable components: the marginal distributions of $\gamma_i^0$ and $\gamma_i^1$, and hence the within-world dependence, are estimable from the observed data, whereas the cross-world correlation $\rho$ is an unidentifiable quantity.  
	\citet{Nevo2022} considered a non-recurrent event setting and specified a correlated gamma distribution for the joint frailty, fixing the location parameters and correlation parameter $\rho$ \textit{a priori} and thereby constraining the model to a relatively restrictive parametric family. In contrast, our method places a flexible nonparametric prior on the joint distribution of $(\gamma_i^0,\gamma_i^1)$ conditional on a fixed $\rho$, specified in Section~\ref{sec:model_specification}.

	Given $\rho$, posterior inference proceeds by sampling $(\boldsymbol{\theta},\boldsymbol{\gamma})$ from $p(\boldsymbol{\theta},\boldsymbol{\gamma}\mid \mathbf{O})
	\propto
	p(\boldsymbol{\theta},\boldsymbol{\gamma}\mid \rho)
	\prod_{i=1}^n p\!\left(U_i,\mathbf{Y}_i \mid Z_i,\mathbf{X}_i,\boldsymbol{\theta},\gamma_i^{Z_i}\right)$ 
	and imputing missing potential outcomes $(U_i^{1-Z_i},\mathbf{Y}_i^{1-Z_i})$ from their posterior predictive distribution.
	This yields posterior draws of the always-survivor indicators and the causal estimands defined in Section~\ref{sec:estimands}.
	Although we denote $\mathbf{Y}_i^z=(Y_{i1}^z,Y_{i2}^z,\ldots)$ as an infinite sequence, computation only requires simulating the latent schedule up to the largest horizon $r$ used to define the principal strata and estimands (and for $t \leq r$).
	Theorem \ref{thm:identification_mu} states that \eqref{eq:estimand_1} is identifiable up to the frailty $\boldsymbol{\gamma}$.

	\begin{theorem}
		\label{thm:identification_mu}
		Under Assumptions \ref{asmp:consistency} -- \ref{asmp:indep_given_fraity}, \eqref{eq:estimand_1} is nonparametrically identified up to distribution of the frailty $\boldsymbol{\gamma}$ as follows:
		\begin{equation}
			\label{eq:identification}
			\begin{split}
				\mu^z\qty(t; r) &= \frac{\int_{\mathcal{X}}\int_{\Gamma} \kappa_{t, r}(z,\mathbf{x},\gamma^z) \eta_{r}(1, \mathbf{x}, \gamma^1) \eta_{r}(0, \mathbf{x}, \gamma^0)  f_{\boldsymbol{\gamma}}(\boldsymbol{\gamma})f_{\mathbf{X}}(\mathbf{x}) d\boldsymbol{\gamma} d\mathbf{x}}
				{\int_{\mathcal{X}}\int_{\Gamma} \eta_{r}(1, \mathbf{x}, \gamma^1) \eta_{r}(0, \mathbf{x}, \gamma^0)  f_{\boldsymbol{\gamma}}(\boldsymbol{\gamma})f_{\mathbf{X}}(\mathbf{x}) d\boldsymbol{\gamma}d\mathbf{x}}, 
			\end{split}
		\end{equation}
		where $\Gamma$ is the support of $\boldsymbol{\gamma}$, $\kappa_{t, r}(z,\mathbf{x},\gamma^z)= \E[N_i(t) \mid Z_i=z, D_i > r, \mathbf{X}_i=\mathbf{x},\gamma^z ]$ is the rate of event occurrence among the observed survivors within each treatment group,  
		and $\eta_{r}(z,\mathbf{x},\gamma^z) = \pr(D_i > r \mid Z_i=z, \mathbf{X}_i=\mathbf{x}, \gamma^z)$ is the conditional survival function within each treatment group.
	\end{theorem}
	\noindent
	The proof is provided in Supplementary Material Section B. The models $\eta_{r}(z,\mathbf{x},\boldsymbol{\gamma})$ and $\kappa_{t, r}(z,\mathbf{x},\boldsymbol{\gamma})$ are induced by the models for $U_i$ and $\mathbf{Y}_i$, respctively. 
	We detail the specification of the working models in Section \ref{sec:BNPmodel}.
	To our knowledge, this identification result for the recurrent event estimand is new and has not been established in the existing literature, including the related work of \citet{Lyu2023}. Theorem~\ref{thm:identification_mu} motivates a new g-computation strategy for estimating the target estimand based on observed data. 
	The distribution of the frailty $f_{\boldsymbol{\gamma}}$ includes the cross-world correlation parameter $\rho$, which cannot be learned from the observed data.  
	Because $\rho$ is non-identifiable, the effect is only partially identified.  Accordingly, we treat $\rho$ as a sensitivity parameter and examine how posterior estimates of the causal estimands change across a range of plausible values, illustrated in Section~\ref{sec:analysis}.

	\begin{remark}[Frailty vs. copula for cross-world dependence in the recurrent-event setting]
		Another potential strategy for cross-world modeling is the copula-based approach of \citet{Xu2022}. In their time-to-event setting, the relevant within-world dependence is between two single event times and can be modeled using a bivariate kernel. By contrast, in our recurrent-event setting, the relevant within-world dependence is between an entire recurrent-event process and a terminal event, making the modeling task fundamentally more difficult because one must characterize the association between a stochastic event-history process and a survival outcome, rather than between two scalar random variables. This added complexity motivates our use of a frailty-based formulation, which offers a natural and tractable way to accommodate such dependence (see Section \ref{sec:model_specification} for details).
	\end{remark}

	\subsection{BNP for recurrent events truncated by a terminal event}
	\label{sec:BNPmodel}
	\subsubsection{Enriched dependent Dirichlet process}
	Although Bayesian causal inference based on Theorem~\ref{thm:identification_mu} can proceed with parametric models for recurrent events and the terminal event, the resulting causal estimators may be prone to model misspecification bias. To mitigate such bias, flexible model specifications, and particularly, Bayesian nonparametric (BNP) methods represent an attractive approach that can better adapt to a wider class of data.

	A popular BNP approach for recurrent processes with a terminal event is jointly modeling the gap times of recurrent and terminal events using a Dirichlet process (DP) mixture model.
	While this is a simple approach adopted in the literature for the joint associational analysis of recurrent events with a terminal event \citep{Paulon2020, Xu2022, Tian2024}, a potential limitation is that the latent random partition induced by the DP will be overwhelmingly determined by the recurrent events rather than the terminal event as the number of recurrences grows, leading to unreliable within-cluster predictions of the survival time, with higher posterior variance due to unnecessarily small sample sizes.
	Additionally, existing approaches use the joint DP primarily to model a frailty term, but it is somewhat counterintuitive that the random partition, which yields the subject-specific frailty, is driven largely by numerous recurrent events rather than the subject-specific survival event.
	More technical discussions on this partitioning issue are provided in  Supplementary Material A.

	We address these issues by introducing a nested structure into the DP prior tailored for the analysis of recurrent events with semi-competing risks.
	The key idea is to introduce a nested partition for the unknown random joint probability measure $\mathbf{P}$. In particular, we consider the random marginal $\mathbf{P}_{\phi}$ and the random conditional $\mathbf{P}_{\theta \mid \phi}$ to obtain the desired clustering structure. Then, the nested prior is defined as $\mathbf{P}_{\phi} \sim \mathrm{DP}(\alpha_{\phi}, P_{0\phi} )$, $\mathbf{P}_{\theta \mid \phi} \sim \mathrm{DP}(\alpha_{\theta}(\phi)$, $P_{0\theta \mid \phi}(\cdot \mid \phi) )$ for all $ \phi \in \Phi$,
	where  $\mathbf{P}_{\theta \mid \phi}$ for $\phi \in \Phi$ are  independent of $\mathbf{P}_{\phi}$. These assumptions induce a prior for the random joint distribution $\mathbf{P}$ through the joint law of the marginal and conditionals and the mapping $(P_{\phi}, P_{\theta \mid \phi}) \to \int P_{\theta \mid \phi}(\cdot \mid \phi ) dP_{\phi}$. Then, the prior is parameterized by the base measure $P_0$, defined by $P_0(A \times B) = \int_A P_{\theta \mid \phi}(B \mid \phi ) dP_{\phi}$. This nested prior is called the \emph{Enriched Dirichlet process (EDP)} prior \citep{Wade2011}. 
	
	One core assumption of the EDP is that the distributions are exchangeable at both survival and recurrent event levels. However, in practice, we often have access to subject-specific covariates $\mathbf{X}_i$, which could potentially include time-indexed covariates $\mathbf{V}_{ij}$ at the $j$-th event (e.g., event number). These variables provide useful information in characterizing the distributions at both levels, and we hence further incorporate the dependence of the covariates on the EDP through the use of the dependent Dirichlet process (DDP) \citep{Maceachern1999} to relax the exchangeability assumption. We refer to the resulting prior as the \emph{enriched dependent Dirichlet process (EDDP)}. The enriched dependent Dirichlet process mixture (EDDPM) uses the EDDP as a prior for the mixing distribution.  
	The key idea behind the EDDP is to define a set of random measures that are marginally EDP-distributed for every possible combination of covariates $\mathbf{x}$ and $\mathbf{v}$. Using a square-breaking construction of the EDP \citep{Wade2011}, 
	the density associated with the EDDP is
	\begin{equation}
		\label{eq:square_breaking}
		f\qty(u, \overline{y} \mid P) 
		= \sum_{k=1}^{\infty}\sum_{l=1}^{\infty} w_k^{\phi}(\mathbf{x})w_{l \mid k}^{\theta}(\mathbf{x},\mathbf{v})K_u\qty(u \mid \overline{y}, \boldsymbol{\phi}_k(\mathbf{x}) )  K_y\qty(\overline{y} \mid  \boldsymbol{\theta}_{l \mid k}(\mathbf{x},\mathbf{v})), 
	\end{equation}
	with some appropriate kernels $K_u$ and $K_y$.
	We introduce the covariate dependence only through atoms by letting $w_k^{\phi}(\mathbf{x})=w_k^{\phi}$ and $w_{l \mid k}^{\theta}(\mathbf{x},\mathbf{v})=w_{l \mid k}^{\theta}$,
	where 
	$w_k^{\phi} = v_k^{\phi} \prod_{l<k}(1-v_l^{\phi})$, 
	$v_k^{\phi} \sim \mathrm{Beta}(1,\alpha_{\phi})$ with $w_1^{\phi}=v_1^{\phi}$, 
	$w_{l \mid k}^{\theta} = v_{l \mid k}^{\theta} \prod_{j<l}(1-v_{j \mid k}^{\theta})$, 
	$v_{l \mid k}^{\theta} \sim \mathrm{Beta}(1,\alpha_{\theta}(\phi))$  with $w_{1 \mid k}^{\theta}=v_{1 \mid k}^{\theta}$ for each $k$, and 
	$\boldsymbol{\phi}_k(\mathbf{x}) \sim P_{0\phi}^{\mathbf{x}}$, 
	$\boldsymbol{\theta}_l(\mathbf{x},\mathbf{v}) \sim P_{0\theta \mid \phi}^{\mathbf{x},\mathbf{v}}$. $\boldsymbol{\phi}_k(\mathbf{x})$ and $\boldsymbol{\theta}_{l \mid k}(\mathbf{x},\mathbf{v})$ are stochastic processes drawn independently from marginal distributions $P_{\phi}^{\mathbf{x}}$ and $P_{0\theta \mid \phi}^{\mathbf{x},\mathbf{v}}(\cdot \mid \phi)$,  indexed by the covariates $\mathbf{x}$ and $\mathbf{v}$. This construction corresponds to the \emph{single-weights} DDP model \citep{Maceachern1999, DeIorio2009}, where the atom processes are indexed by covariates, but the weights are independent of the covariates.

	\subsubsection{Model specifications}
	\label{sec:model_specification}
	For $i=1,\ldots,n$ and $j=1,\ldots,N_i$, the log-scale survival time $U_i$ and $j$-th recurrent gap time $Y_{ij}$ are specified by the following hierarchical model.
	\begin{equation}
		\label{eq:model_specification}
		\begin{split}
			U_i \mid Z_i=z, \mathbf{X}_{i}=\mathbf{x} &\sim K_u(\boldsymbol{\phi}_{i}(\mathbf{x},z)) \equiv \mathrm{N}\qty( (\mathbf{x}^\top, z) \boldsymbol{\beta}_{u,i} + \gamma_i^z, \tau_i^2 ),\\
			Y_{ij} \mid Z_i=z, \mathbf{X}_{i}=\mathbf{x}, \mathbf{V}_{ij}=\mathbf{v}
			&\sim K_y(\boldsymbol{\theta}_{ij}(\mathbf{x},\mathbf{v}, z)) \equiv \mathrm{N}\qty( (\mathbf{x}^\top,\mathbf{v}^\top, z) \boldsymbol{\beta}_{y,ij}  + \psi_{ij}\gamma_i^z, \sigma_{ij}^2),\\
			\boldsymbol{\phi}_{i}(\mathbf{x},z) &\sim P_{\phi}^{\mathbf{x}}, ~~ P_{\phi}^{\mathbf{x}} \sim \text{DP}(\alpha_{\phi}, P_{0\phi}^{\mathbf{x}}) \\
			\boldsymbol{\theta}_{ij}(\mathbf{x},\mathbf{v},z)  &\sim P_{\theta \mid \phi}^{\mathbf{x},\mathbf{v}}, ~~ P_{\theta \mid \phi}^{\mathbf{x},\mathbf{v}}
			\sim \text{DP}(\alpha_{\theta \mid \phi}(\phi), P_{0\theta \mid \phi}^{\mathbf{x},\mathbf{v}}(\cdot \mid \phi)),
		\end{split}
	\end{equation}
	where the atom processes  are expressed by simple linear models and variance parameters: $\boldsymbol{\phi}_{i}(\mathbf{x},z) = \qty((\mathbf{x}^\top, z) \boldsymbol{\beta}_{u,i}, \gamma_i^z, \tau_i )$ and
	$\boldsymbol{\theta}_{ij}(\mathbf{x},\mathbf{v},z) = \qty( (\mathbf{x}^\top,\mathbf{v}^\top, z) \boldsymbol{\beta}_{y,ij}, \psi_{ij},  \sigma_{ij}, \gamma_i^z)$. 
	Although we write unit- and gap-indexed latent parameters $(\boldsymbol{\phi}_i,\boldsymbol{\theta}_{ij})$, under the enriched Dirichlet process prior, these parameters take on only finitely many distinct values almost surely.
	Thus, the model induces a data-adaptive clustering in which many units (and many gaps) share the same parameter values; the effective number of free parameters is the number of occupied clusters rather than $n$.
	Additionally, we assume $\alpha_{\phi} \sim \mathrm{Ga}(a_{\alpha_{\phi}}, b_{\alpha_{\phi}})$, 
	$\alpha_{\theta \mid \phi}(\phi)=\alpha_{\theta \mid \phi} \sim \mathrm{Ga}(a_{\alpha_{\theta \mid \phi}}, b_{\alpha_{\theta \mid \phi}})$, 
	and the marginal base measures are $ P_{0\phi}^{\mathbf{x}} = 
	\mathrm{MVN}(\boldsymbol{\mu}_{\boldsymbol{\beta}_u}, \boldsymbol{\Sigma}_{\boldsymbol{\beta}_u}) 
	\times \mathrm{MVN}(\boldsymbol{\mu}_{\boldsymbol{\gamma}}, \boldsymbol{\Sigma}_{\boldsymbol{\gamma}})
	\times \mathrm{IG}(a_{\tau}, b_{\tau})$, where $\boldsymbol{\Sigma}_{\boldsymbol{\gamma}}=\left(\begin{smallmatrix}
		\sigma_{\gamma_0}^2 & \rho\sigma_{\gamma_0}\sigma_{\gamma_1} \\
		\rho\sigma_{\gamma_0}\sigma_{\gamma_1} & \sigma_{\gamma_1}^2
	\end{smallmatrix}\right)$ with a given correlation parameter $\rho$, and 
	$P_{0\theta \mid \phi}^{\mathbf{x},\mathbf{v}}(\cdot \mid \phi) = \mathrm{MVN}(\boldsymbol{\mu}_{\boldsymbol{\beta}_y}, \boldsymbol{\Sigma}_{\boldsymbol{\beta}_y})
	\times\mathrm{N}(\mu_{\psi}, \sigma^2_{\psi})
	\times\mathrm{IG}(a_{\sigma}, b_{\sigma}) \times\delta_{\boldsymbol{\gamma}}$, 
	where $\delta_{\boldsymbol{\gamma}}$ denotes a Dirac measure at the frailty component $\boldsymbol{\gamma}$ contained in $\phi$,
	so that the same subject-level frailty is propagated from the terminal-time kernel into the recurrent-gap kernel within each top-level $\phi$-cluster.
	We will discuss specific choices of the hyperparameters of each prior in the simulation studies and empirical analyses.

	The hierarchical model \eqref{eq:model_specification} is equivalent to jointly modeling the observed recurrent gap times and survival time for each unit through the infinite mixture of the log-normal kernels. 
	It is worth mentioning that many existing approaches to the analysis of recurrent and survival events in non-causal analyses only place a DP prior on the random frailty of a linear model and/or the residuals. While their model relaxes the distributional assumption on the random effects in the linear model, it still makes strong structural assumptions about how the parametric fixed effects are correlated with the outcome (i.e., linearity assumption). In contrast, our model is intrinsically \emph{functional}, placing EDDPM priors on the functional space of the survival and recurrent event models.
	
	Another salient feature of the proposed EDDPM for recurrent-terminal event data is its treatment of the subject-specific frailty vector 
	$\boldsymbol{\gamma}_i=(\gamma_i^0,\gamma_i^1)$.
	We first sample $\boldsymbol{\gamma}_i$ from the first-level marginal base measure $P_{0\phi}^{\mathbf{x}}$, which governs the terminal-event distribution, and then propagate this same value through the second-level conditional base measure  
	$P_{0\theta\mid\phi}^{\mathbf{x},\mathbf{v}}(\,\cdot\mid\phi) \propto\delta_{\boldsymbol{\gamma}_i},$ so that the nested EDDP prior carries identical frailty information into the recurrent gap-time model.
	The frailty $\boldsymbol{\gamma}_i$ captures unobserved heterogeneity for the terminal event, while the parameters $\psi_{ij}$ determine how this heterogeneity modifies each recurrent gap time for subject $i$.
	This nested construction avoids the unfavorable random-partition behavior highlighted in Section~\ref{sec:BNPmodel} (see also Section \ref{sec:random_partition}).
	Conditional on $\boldsymbol{\gamma}_i$ we assume independence between the terminal event and the recurrent events, a standard simplifying device in descriptive analyses of such data \citep{Paulon2020, Xu2021, Tian2024}.
	
	We parameterize cross-world dependence through the frailty-level correlation
	$\rho=\Cor(\gamma_i^0,\gamma_i^1)$ by placing a bivariate distribution with correlation $\rho$
	in the base measure for the frailty component of the top-level EDDP atom.
	Concretely, the $\boldsymbol{\gamma}$-margin of the top-level base measure is
	$\operatorname{MVN}\!\bigl(\boldsymbol{\mu}_{\boldsymbol{\gamma}},\boldsymbol{\Sigma}_{\boldsymbol{\gamma}}\bigr)$, where $\boldsymbol{\Sigma}_{\boldsymbol{\gamma}}=\left(\begin{smallmatrix}
		\sigma_{\gamma_0}^2 & \rho\sigma_{\gamma_0}\sigma_{\gamma_1} \\
		\rho\sigma_{\gamma_0}\sigma_{\gamma_1} & \sigma_{\gamma_1}^2
	\end{smallmatrix}\right)$.
	Because the EDDP is centered on its base measure, the induced prior predictive marginal law of a randomly
	selected subject's frailty vector $\boldsymbol{\gamma}_i$ equals this $\boldsymbol{\gamma}$-margin, and therefore
	$\Cor(\gamma_i^0,\gamma_i^1)=\rho$ under the prior.
	In our enriched construction, $\boldsymbol{\gamma}_i$ is sampled at the top level (through the $\phi$-atoms governing
	the terminal event) and then propagated to the recurrent-event kernel via the conditional base measure
	$P_{0\theta\mid\phi}^{\mathbf{x},\mathbf{v}}(\cdot\mid \phi)\propto \delta_{\boldsymbol{\gamma}_i}$, so that the same
	subject-level frailty enters both the terminal- and recurrent-event components within each top-level cluster.
	
	Inference is performed conditional on a fixed $\rho$, which is not identified from the observed data because
	$(\gamma_i^0,\gamma_i^1)$ are never jointly observed.
	Accordingly, we treat $\rho$ as a sensitivity parameter and examine the robustness of posterior inferences for the causal
	estimands over a plausible grid of $\rho$ values.
	We emphasize that $\rho$ governs cross-world association at the frailty level, inducing the dependence structure between potential outcomes, but the induced association (e.g., $U_i^0$ and $U_i^1$) may differ due to additional outcome-specific variability and model components beyond $\boldsymbol{\gamma}_i$.
	Overall, this formulation allows for a flexible, nonparametric characterization of the survival event distribution, while still permitting subject-level random effects $\boldsymbol{\gamma}_i$ that may be correlated with recurrent gap times through $\psi_{ij} \boldsymbol{\gamma}_i$.

	\subsubsection{Random partition properties}
	\label{sec:random_partition}
	When applied to recurrent event data with a terminal event, the standard DP induces a random partition that is predominantly determined by the recurrent events rather than the terminal event. We provide detailed discussions on this unfavorable property of the DP-induced partition in Supplementary Material Section A.2.
	This section discusses how the random partition induced by our proposed prior addresses this limitation.  
	Define $\mathcal{P}_n = (\mathcal{P}_{n,u}, \mathcal{P}_{n,y}),$
	where $\mathcal{P}_{n,u} = (s_{u,1}, \dots, s_{u,n})$ and $\mathcal{P}_{n,y} = (s_{y,1}, \dots, s_{y,n})$ denote random partitions defined by cluster allocation labels. Specifically, $s_{u,i} = j$ if $\phi_i = \phi^*_j$ is the $j$-th distinct value of the atom $\phi$, and $s_{y,i} = l$ if $\theta_i = \theta^*_{l \mid j}$ is the $l$-th distinct value of the atom $\theta$ within the $j$-th $\phi$-cluster.
	For each $\phi$-cluster $j$, let $\mathcal{S}_j = \{ i : s_{u,i} = j \},$
	so $n_j = |\mathcal{S}_j|$ is the number of individuals in the $j$-th $\phi$-cluster. Denote the total number of $\phi$-clusters by $M_n$. Within each $\phi$-cluster $j$, define $\mathcal{S}_{l \mid j} = \{ i : s_{u,i} = j,  s_{y,i} = l \}$, 
	so $n_{l \mid j} = |\mathcal{S}_{l \mid j}|$ is the size of the $l$-th $\theta$-subcluster, and let $M_{n,j}$ be the number of $\theta$-subclusters within cluster $j$. The unique cluster-specific parameters are then $\phi^* = (\phi^*_j)_{j=1}^{M_n}$ and $\theta^* = \qty(\theta^*_{1 \mid j}, \dots, \theta^*_{M_{n,j} \mid j})_{j=1}^{M_n}$.
	Finally, define $\mathcal{P}_{n_j,y} = (s_{y,i}: i \in \mathcal{S}_j)$, $U_j^* = \{U_i : i \in \mathcal{S}_j\}$, $\overline{Y}_j^* = \{\overline{Y}_i : i \in \mathcal{S}_j\}$, $\overline{Y}_{l \mid j}^* = \{\overline{Y}_i : i \in \mathcal{S}_{l \mid j}\}$.
	
	Consider the joint posterior distribution of the partition and the cluster parameters:
	\begin{equation*}
		\begin{split}
			p(\mathcal{P}_n, \phi^*, \theta^* &\mid \overline{Y}_{1:n}, U_{1:n}) \\
			\propto &
			p(\mathcal{P}_n)
			\prod_{j=1}^{M_n} p_{0\phi}(\phi_j^*) 
			\prod_{l=1}^{M_{n,j}} p_{0\theta}(\theta_{l \mid j}^*) 
			\prod_{j=1}^{M_n} \prod_{i \in \mathcal{S}_j}
			K\left(U_i \mid \overline{Y}_i, \phi_j^*\right)
			\prod_{l=1}^{M_{n,j}} \prod_{i \in \mathcal{S}_{l \mid j}}
			K\left(\overline{Y}_i \mid \theta_{l \mid j}^*\right).
		\end{split}
	\end{equation*}
	The posterior distributions of the cluster-specific parameters are 
	\begin{equation}
		\label{eq:post_parameters_EDP}
		\begin{split}
			p\left(\phi_j^* \mid \mathcal{P}_n, \overline{Y}_{1:n}, U_{1:n}\right)
			&\propto 
			p_{0\phi}\left(\phi_j^*\right)
			\prod_{i \in \mathcal{S}_j} 
			K\left(U_i \mid \overline{Y}_i, \phi_j^*\right),\\
			p\left(\theta_{l \mid j}^* \mid \mathcal{P}_n, \overline{Y}_{1:n}, U_{1:n}\right)
			&\propto 
			p_{0\theta}\left(\theta_{l \mid j}^*\right)
			\prod_{i \in \mathcal{S}_{l \mid j}} 
			K\left(\overline{Y}_i \mid \theta_{l \mid j}^*\right).
		\end{split}
	\end{equation}
	\noindent
	The EDP random partition is further characterized by the following proposition.
	\begin{proposition}
		\label{prop:1}
		Let $h_u\left(U_j^* \mid \overline{Y}_j^*\right) 
		= \int_{\Phi} \prod_{i \in \mathcal{S}_j} K\left(U_i \mid \overline{Y}_i, \phi\right)
		dP_{0\phi}(\phi)$ and 
		$h_y\left(\overline{Y}_{l \mid j}^*\right) 
		= \int_{\Theta} \prod_{i \in \mathcal{S}_{l \mid j}} K\left(\overline{Y}_i \mid \theta\right)
		dP_{0\theta}(\theta)$.
		The posterior of the random partition of the EDP model is
		\begin{equation}
			\label{eq:posterior_partition_EDP}
			\begin{split}
				&p(\mathcal{P}_n \mid \overline{Y}_{1:n}, U_{1:n}) \\
				&\propto 
				\alpha_{\phi}^{M_n} \prod_{j=1}^{M_n} \int_{\Phi} \alpha_{\theta}(\phi)^{M_{n,j}} \frac{\Gamma(\alpha_{\theta}(\phi)) \Gamma(n_j)}{\Gamma(\alpha_{\theta}(\phi) + n_j)} dP_{0\phi}(\phi)
				h_u\left(U_j^* \mid \overline{Y}_j^* \right) 
				\prod_{l=1}^{M_{n,j}}\Gamma(n_{l \mid j})h_y\left(\overline{Y}_{l \mid j}^*\right).
			\end{split}
		\end{equation}
	\end{proposition}
	\noindent
	The proof is provided in Supplementary Material Section B. 
	Moreover, by marginalizing over all possible $\theta$-subpartitions for each $\phi$-cluster, one obtains the posterior for $\mathcal{P}_{n,u}$:
	\begin{equation}
		\label{eq:posterior_partition_EDP_u}
		\begin{split}
			p(\mathcal{P}_{n,u} &\mid \overline{Y}_{1:n}, U_{1:n}) 
			\propto 
			\alpha_{\phi}^{M_n} \prod_{j=1}^{M_n} h_u\left(U_j^* \mid \overline{Y}_j^* \right)  \\
			& \times \sum_{\mathcal{P}_{n_j,y} \in \Pi_{n_j}} \int_{\Phi} \alpha_{\theta}(\phi)^{M_{n,j}} \frac{\Gamma(\alpha_{\theta}(\phi)) \Gamma(n_j)}{\Gamma(\alpha_{\theta}(\phi) + n_j)} dP_{0\phi}(\phi)
			\prod_{l=1}^{M_{n,j}}\Gamma(n_{l \mid j})h_y\left(\overline{Y}_{l \mid j}^*\right),
		\end{split}
	\end{equation}
	where $\Pi_{n_j}$ denotes the set of all possible partitions of $n_j$ integers. 
	The posterior \eqref{eq:posterior_partition_EDP_u} shows that $\phi$-clusters favored under the EDP are those in which individuals share a similar survival-recurrence relationship (reflected in $h_u$), while finer distinctions among recurrent outcomes (captured by a mixture of the kernel of $h_y$) appear as nested $\theta$-subclusters.

	The posterior distributions \eqref{eq:posterior_partition_EDP} and \eqref{eq:posterior_partition_EDP_u} both reflect the desirable random-partition structure of the EDP. In particular, the nested framework separates the likelihood contributions from $U_j^*$ (the survival component for individuals in the $j$-th $\phi$-cluster) and $\overline{Y}_{l \mid j}^*$ (the recurrent events for individuals in the $l$-th $\theta$-subcluster within cluster $j$). Hence, even when the data indicate that many $\theta$-subclusters are needed to capture fine differences in the recurrence process, these subclusters all remain nested under the coarser $\phi$-cluster. As a result, the number of top-level $\phi$-clusters, $M_n$, tends to remain modest, yielding a larger effective sample size $n_j$ within each $\phi$-cluster. From \eqref{eq:post_parameters_EDP}, we see that $\phi_j^*$ is updated using all $U_i$ and $\overline{Y}_i$ data in $\mathcal{S}_j$, regardless of how the cluster is further subdivided. This more substantial pool of observations stabilizes posterior estimates of $\phi_j^*$.

	\subsubsection{Posterior inference}
	We develop a fully tractable Gibbs sampling algorithm for the posterior inference.
	First, the observed-data likelihood 
	is given by:
	\begin{equation*}
		\begin{split}
			\mathcal{L}_{\mathrm{obs}} 
			=& \prod_{i=1}^{n} f^{1-\delta^C_i}\left(U_i \mid Z_i, \mathbf{X}_{i}, \boldsymbol{\beta}_{u,i}, \gamma_i^{Z_i}, \tau_i\right)
			S_U^{\delta^C_i}\left(\log(\mathcal{T}_i) \mid Z_i, \mathbf{X}_{i}, \boldsymbol{\beta}_{u,i}, \gamma_i^{Z_i}, \tau_i\right) \\
			& \times 
			\prod_{j=1}^{N_i} f\left(Y_{ij} \mid Z_i,\mathbf{X}_{i},\mathbf{V}_{ij}, \boldsymbol{\beta}_{y,ij}, \psi_{ij}, \gamma_i^{Z_i}, \sigma_{ij} \right)  \\
			& \times ~ S_Y\left(\log(\mathcal{T}_i - T_{iN_i}) \mid Z_i,\mathbf{X}_{i},\mathbf{V}_{i(N_i+1)}, \boldsymbol{\beta}_{y,i(N_i+1)}, \psi_{i(N_i+1)}, \gamma_i^{Z_i}, \sigma_{i(N_i+1)} \right),
		\end{split}
	\end{equation*}
	where $S_Y$ and $S_U$ denote the survival functions for $Y$ and $U$, respectively. 
	To facilitate the posterior inference based on the data augmentation, we consider the following complete data likelihood with truncated outcomes imputed.
	\begin{equation*}
		\begin{split}
			\mathcal{L}_{\mathrm{comp}} 
			=& \prod_{i=1}^{n} f^{1-\delta^C_i}\left(U_i \mid Z_i,\mathbf{X}_{i}, \boldsymbol{\beta}_{u,i}, \gamma_i^{Z_i}, \tau_i\right)
			f^{\delta^C_i}\left(U^{*}_i \mid Z_i,\mathbf{X}_{i}, \boldsymbol{\beta}_{u,i}, \gamma_i^{Z_i}, \tau_i\right)\\
			& \times  
			\prod_{j=1}^{N_i} f\qty(Y_{ij} \mid Z_i,\mathbf{X}_{i},\mathbf{V}_{ij}, \boldsymbol{\beta}_{y,ij}, \psi_{ij}, \gamma_i^{Z_i}, \sigma_{ij}) \\
			& \times f\qty(Y^{*}_{i(N_i+1)} \mid Z_i,\mathbf{X}_{i},\mathbf{V}_{i(N_i+1)}, \boldsymbol{\beta}_{y,i(N_i+1)}, \psi_{i(N_i+1)}, \gamma_i^{Z_i}, \sigma_{i(N_i+1)}),
		\end{split}
	\end{equation*}
	where $Y^{*}$ and $U^{*}$ represent the imputed values of the gap time and survival time, respectively. 
	This complete likelihood admits the standard posterior sampling technique based on the densities.
	Specifically, we employ an approximate blocked Gibbs sampler based on a two-level truncation of the square-breaking representation of the EDP proposed by \citet{Burns2023}.
	In this algorithm, we first select conservative upper bounds on the number of latent classes of the square-breaking representation of \eqref{eq:square_breaking} as follows.
	\begin{align*}
		f\qty(u, \overline{y} \mid P) 
		&= \sum_{k=1}^{K}\sum_{l=1}^{L} w_k^{\phi}w_{l \mid k}^{\theta}\mathrm{N}\qty(u \mid (\mathbf{x}, z)^\top \boldsymbol{\beta}_{u,k} + \gamma_k^z, \tau_k^2 )  \mathrm{N}\qty(\overline{y} \mid (\mathbf{x},\mathbf{v}, z)^\top \boldsymbol{\beta}_{y,l \mid k} + \psi_{l \mid k}\gamma_k^z , \sigma_{l \mid k}^2) 
	\end{align*}
	\noindent
	Let $G_i \in \{1,...,K\}$ and $H_{ij} \in \{1,...,L\}$ denote the latent cluster indicators for individual $i=1,\ldots,n$ and time $j=1,\ldots,N_i$. Our data augmentation algorithm further imputes these latent indicators in each iteration to facilitate the posterior updates of model parameters.
	Specifically, we specify Multinomial distributions $G_i \sim \mathrm{MN}(\mathbf{w}^{\phi})$ on $G_i$ and $H_{ij} \sim \mathrm{MN}(\mathbf{w}^{\theta}_k)$ on $H_{ij}$,  where $\mathbf{w}^{\phi} = (w^{\phi}_{1},\ldots,w^{\phi}_K)^\top$  and  $\mathbf{w}^{\theta}_k = (w^{\theta}_{1 \mid k},\ldots,w^{\theta}_{L \mid k
	})^\top$ contains the weights from the EDDP. 
	\citet{Burns2023} demonstrated that an accurate approximation to the exact EDP is obtained as long as the truncation bound is sufficiently large. 
	To ensure this, we ran several MCMC iterations with different values of $K$ and $L$ and increased them if all clusters were occupied. 
	Overall, our algorithm iterates between drawing from the conditional distributions of censored outcomes, latent cluster indicators, and model parameters.
	
	The essential Gibbs sampler is outlined as follows: (i) given all model parameters, $G_i$ and $H_i$, sample $Y^{*}_{i(N_i+1)}$ and $U^{*}_i$, (ii)
	given all model parameters, $Y^{*}_{i(N_i+1)}$ and $U^{*}_i$, sample $G_i$ and $H_i$, (iii) given $Y^{*}_{i(N_i+1)}$, $U^{*}_i$, $G_i$ and $H_i$, sample all model parameters, and (iv) compute the estimands based on Theorem \ref{thm:identification_mu}.
	Specifically,  when imputing $(N_i+1)$-th gap time $Y^{*}_{i(N_i+1)}$ in the first step, we draw from its conditional predictive distribution $p(Y \mid Y > \log(\mathcal{T}_i - T_{iN_i}), \bullet)$, where ``$\bullet$'' denotes the rest of relevant variables, as the final gap time is always censored at $\mathcal{T}_i - T_{iN_i}$, the actual gap time should be greater than the censored gap time.
	Similarly, for the unit with $\delta_i^C=1$,  its survival time $U_i$ from the posterior predictive distribution enables us to determine if the unit belongs to $\mathcal{AS}(r)$. For the imputation of $U_i$ with $\delta_i^C=1$, we also draw from its conditional predictive distribution $p(U \mid U > \log(\mathcal{T}_i), \bullet)$ because the observed survival time is right-censored and the realized $U_i$ must exceed the censoring bound.
	When computing the estimands in step (iv), we approximate the integrals in~\eqref{eq:identification} by Monte Carlo simulation.
	For example, for a fixed arm $z\in\{0,1\}$, the numerator of~\eqref{eq:identification} for $\mu^z(t;r)$ is approximated by
	$\frac{1}{n}\sum_{i=1}^{n}\kappa_{t,r}\!\left(z,X_i,\boldsymbol{\gamma}_i\right)\,
	\eta_{r}\!\left(1,X_i,\boldsymbol{\gamma}_i\right)\,
	\eta_{r}\!\left(0,X_i,\boldsymbol{\gamma}_i\right),$
	with an analogous sample-average approximation for the denominator without the need to additionally model the distribution of baseline covariates.
	This is a conventional approximation in the Bayesian paradigm \citep{Li_Ding2023}.  See Supplementary Material Section C for full details of the algorithm.
	The initial parameter values were randomly drawn from the prior distributions, and the posterior samples were obtained by running a chain for $10000$ MCMC iterations after an initial $40000$ burn-in iterations. 
	Convergence was monitored by the trace plots, confirming that the chains had reached stationarity with good mixing.

	\section{Simulation Studies}
	\label{sec:simulation}
	We examine the performance of the proposed methods through simulation studies. Specifically, here we evaluate the frequentist properties of the proposed EDDP prior for estimating the key estimands $\mu^z(t;r)$ 
	for $z \in \{0,1\}$.
	We simulate $300$ datasets with $n=1000$ individuals and report the bias and root mean square error (RMSE) of a point estimator (posterior mean), as well as the frequentist coverage probability (CP) and average length (AL) of the $95\%$ central credible interval. The dataset is generated from the following mixture models with random seeds: 
	\begin{align*}
		U_{i}^z &\sim \sum_{k=1}^{3} w_k\mathrm{N}(\alpha^u + \mathbf{X}_{i}^\top \phi_{k} + z \beta^u_k + \gamma_i^z, 0.2),\\
		Y_{ij}^z &\sim \sum_{k=1}^{3} \pi_k \mathrm{N}(\alpha^y + \mathbf{X}_{i}^\top \theta_{k} + z \beta^y_k +  \psi \gamma_i^z, 0.2),
	\end{align*}
	where $\mathbf{X}_i \sim \mathrm{MVN}((0,0,0)^\top, I_{3} )$ with $I_{d}$ is the identity matrix of dimension $d$, $Z_i \sim \mathrm{Bern}(0.5)$, 
	$(w_1,w_2,w_3)=(0.3,0.4,0.3)$, 
	$(\pi_1,\pi_2,\pi_3)=(0.3,0.4,0.3)$, 
	$\alpha^u=6.5$, $\alpha^y=5.0$, 
	$\phi_1 = (0.2, 0.15, -0.1)^\top$, $\phi_2=-0.5\phi_1$, $\phi_3=0.3\phi_1$, $\theta_1 = (0.25, -0.10, -0.15)^\top$, $\theta_2=-0.5\theta_1$, $\theta_3=0.3\theta_1$, $(\beta^u_1,\beta^u_2,\beta^u_3)=(1.0,0.5,1.3)$, $(\beta^y_1,\beta^y_2,\beta^y_3)=(0.5,0.25,0.65)$, and $\psi=0.1$. 
	The frailty follows the multivariate log-normal distribution such that $\boldsymbol{\gamma}_i = \exp(\boldsymbol{\gamma}_i')$, where $\boldsymbol{\gamma}_i' \sim \mathrm{MVN}((0,0)^\top, \left(\begin{smallmatrix}
		0.2 & 0.2\rho \\
		0.2\rho & 0.2
	\end{smallmatrix}\right))$ with a correlation parameter $\rho = 0.5$.
	The mixture components $w_k$ and $\pi_k$ take the same weight value for $k=1,2,3$, but the assignment to each distribution component is independent. The observed survival time and recurrent gap times are obtained by transforming the log-scale variables $U_{i}^z$ and $Y_{ij}^z$ into $D_i = \exp(U_{i}^{Z_i})$ and $W_{ij} = \exp(Y_{ij}^{Z_i})$, and $W_{ij}$ is generated recurrently until the sum exceeds the censoring time $\mathcal{T}_i = \min(D_i,C_i)$ where $ C_i \sim \mathrm{Unif}(300,1000)$, where $C_i$ represents the independent censoring time for each individual. The estimands’ true values are approximated by Monte Carlo simulation using large samples.
	
	\begin{table}[!htbp]
		\centering
		\caption{Bias and root mean squared error (RMSE) of point estimates, and coverage probability (CP) and average length (AL) of $95\%$ central credible intervals of $\mu^z(t;r)$ for $(t,r)=(300, 500), (500,500)$.}
		\begin{adjustbox}{width=14.cm}
			\begin{tabular}{cl rrrr rrrr}
				\toprule
				& & \multicolumn{4}{c}{$\mu^0(t;r)$} & \multicolumn{4}{c}{$\mu^1(t;r)$}  \\
				\cmidrule(lr){3-6} \cmidrule(lr){7-10}
				$(t,r)$ & Model & Bias & RMSE & CP & AL & Bias & RMSE & CP & AL \\
				\midrule
				\multirow{4}{*}{$(300,500)$}        & LM    & 0.05 & 0.09 & 0.23 & 0.12 & 0.04 & 0.08 & 0.18 & 0.09 \\
				& DPM   & 0.08 & 0.08 & 0.78 & 0.17 & 0.01 & 0.06 & 0.45 & 0.09 \\
				& DDPM  & 0.03 & 0.06 & 0.83 & 0.12 & 0.00 & 0.04 & 0.74 & 0.09 \\
				& EDDPM & 0.01 & 0.03 & 0.85 & 0.09 & 0.02 & 0.03 & 0.80 & 0.08 \\
				\midrule                            
				\multirow{4}{*}{$(500,500)$}        & LM    & 0.09  & 0.15 & 0.16 & 0.18 & 0.06  & 0.12 & 0.10 & 0.13 \\
				& DPM   & 0.11  & 0.12 & 0.92 & 0.30 & -0.00 & 0.09 & 0.44 & 0.14 \\
				& DDPM  & 0.03  & 0.08 & 0.84 & 0.18 & -0.01 & 0.06 & 0.67 & 0.13 \\
				& EDDPM & -0.01 & 0.05 & 0.87 & 0.14 & 0.01  & 0.04 & 0.87 & 0.12 \\
				\bottomrule
			\end{tabular}
		\end{adjustbox}
		\label{tab:simulation_result_05}
	\end{table}
	
	We compare our proposed EDDPM \eqref{eq:model_specification} with the models commonly used in the recurrent event analysis: linear model (LM) \citep[e.g.,][]{Comment2025, Lyu2023}, Dirichlet process mixture (DPM) \citep[e.g.,][]{Paulon2020, Tian2024}, and dependent Dirichlet process mixture (DDPM) \citep[e.g.,][]{Xu2022}. In DDPM, the nested prior structure is removed from EDDPM. In DPM, covariate dependence is further removed from DDPM, and the frailty and error terms are modeled using DPM, resulting in a model equivalent to that of \citet{Paulon2020}. For LM, we fit a linear mixed model to the log-transformed survival and gap times. The specification includes a subject-level frailty term and assumes normal residual errors, which yields the familiar accelerated failure time (AFT) representation. This setup is close to the models explored by \citet{Comment2025} and \citet{Lyu2023} and is strictly parametric. The only difference is that their proportional hazard model assumes an extreme-value distribution for the error term. We use proper, weakly informative conjugate priors for all parameters. Specifically, we choose the hyperparameters such that $a_{\alpha_{\phi}}=a_{\alpha_{\theta \mid \phi}} =a_{\tau}=a_{\sigma}=2.0$, $b_{\alpha_{\phi}}=b_{\alpha_{\theta \mid \phi}}=b_{\tau}=b_{\sigma}=1.0$, $\mu_{\gamma}=\mu_{\psi}=0.0$, $\sigma^2_{\gamma}=\sigma^2_{\psi}=3.0^2$, $\boldsymbol{\mu}_{\boldsymbol{\beta}_u}$ and $\boldsymbol{\mu}_{\boldsymbol{\beta}_y}$ are zero-vectors of appropriate sizes, and $\boldsymbol{\Sigma}_{\boldsymbol{\beta}_u}$ and $\boldsymbol{\Sigma}_{\boldsymbol{\beta}_y}$ are diagonal matrices with $3.0^2$ on the diagonal elements.

	Table \ref{tab:simulation_result_05} presents the simulation results. 
	Overall, the results consistently demonstrate that our method yields unbiased estimates and has the smallest RMSE across all scenarios, indicating EDDPM's superior accuracy in point estimation. 
	Turning to interval estimation, which is assessed by the coverage probability (CP) and average length (AL) of $95\%$ credible intervals, EDDPM stands out for better CP compared to other methods, closer to the frequentist target $95\%$, and shorter interval lengths (reflecting improved efficiency). The linear model is sensitive to latent treatment heterogeneity within the data-generating processes, resulting in the poorest performance across most metrics. While the standard DP models offer improvement over the parametric model, they fall short of the high standards set by EDDPM. In practice, we recommend evaluating the predictive performance of these models, as illustrated in Section \ref{sec:analysis}.

	\section{Analysis of HF-ACTION Randomized Clinical Trial: Investigating the Effect of Exercise Training in Patients with Chronic Heart Failure}
	\label{sec:analysis}
	
	We analyze the HF-ACTION trial introduced in Section~\ref{sec:motivating_example} using our proposed principal stratification framework. HF-ACTION randomized medically stable outpatients with chronic systolic heart failure to usual care alone versus usual care plus a structured aerobic exercise-training program \citep{OConnor2009}. In our analysis, the recurrent outcome is the sequence of all-cause hospitalizations over follow-up, while death is treated as an absorbing terminal event that stops further observation of subsequent hospitalizations. This setting is typical for treating patients with chronic heart failure as the cumulative burden and tempo of rehospitalizations are central measures of morbidity, yet they are intertwined with survival because longer survival increases time at risk for hospitalization while death truncates the recurrent process.
	
	\subsection{Principal stratification analysis in continuous time}
	
	Our analysis targets causal effects on rehospitalization while holding survivorship fixed by restricting attention to always-survivor principal stratum. Specifically, our primary causal estimand is the survivor-average number of recurrence ratio, 
	$$\mathrm{SANR}(t;r)=\mu^1(t;r)\big/\mu^0(t;r),$$
	and $\mu^z(t;r)=\E\!\left\{N_i^z(t)\mid \mathcal{AS}(r)\right\}$. Here $z=1$ indicates exercise training and $z=0$ indicates usual care. Thus, $\mathrm{SANR}(t;r)<1$ means that among patients who would be alive up to time $r$ regardless of assignment, exercise training reduces the expected cumulative hospitalization burden up to time $t$. The estimand $\mathrm{SANR}(t;r)$ answers a counterfactual question directly relevant to the burden of heart failure management: if we could follow the same set of patients who would remain alive through time $r$ under both strategies, how would prescribing structured exercise change the expected number of hospitalizations accumulated by time $t$?
	Having defined the causal estimand of interest, we turn to the modeling strategies used to estimate it and assess their empirical fit. Specifically, we compare the four models considered in Section \ref{sec:simulation} using the log pseudo marginal likelihood (LPML) \citep{Geisser1979}.
	The LPML is a Bayesian model-fit criterion derived from leave-one-out (LOO) predictive assessments of the data. 
	The LPML values for LM, DPM, DDPM, and EDDPM are $-20913.96$, $-17689.37$, $-15712.22$, and $-15458.95$, respectively. Since a higher LPML indicates a better predictive fit in terms of LOO predictive densities, the EDDPM emerges as the superior model among the four models. We therefore present analysis results under EDDPM \eqref{eq:model_specification} with weakly informative conjugate priors in Section \ref{sec:simulation}. Throughout this subsection, we report results for the cross-world dependence sensitivity parameter $\rho=0.5$. This is a neutral assumption in the absence of external knowledge and sensitivity analysis will be reported in Section \ref{sec:SA}.
	
\begin{figure}
		\centering
		\includegraphics[width=7cm]{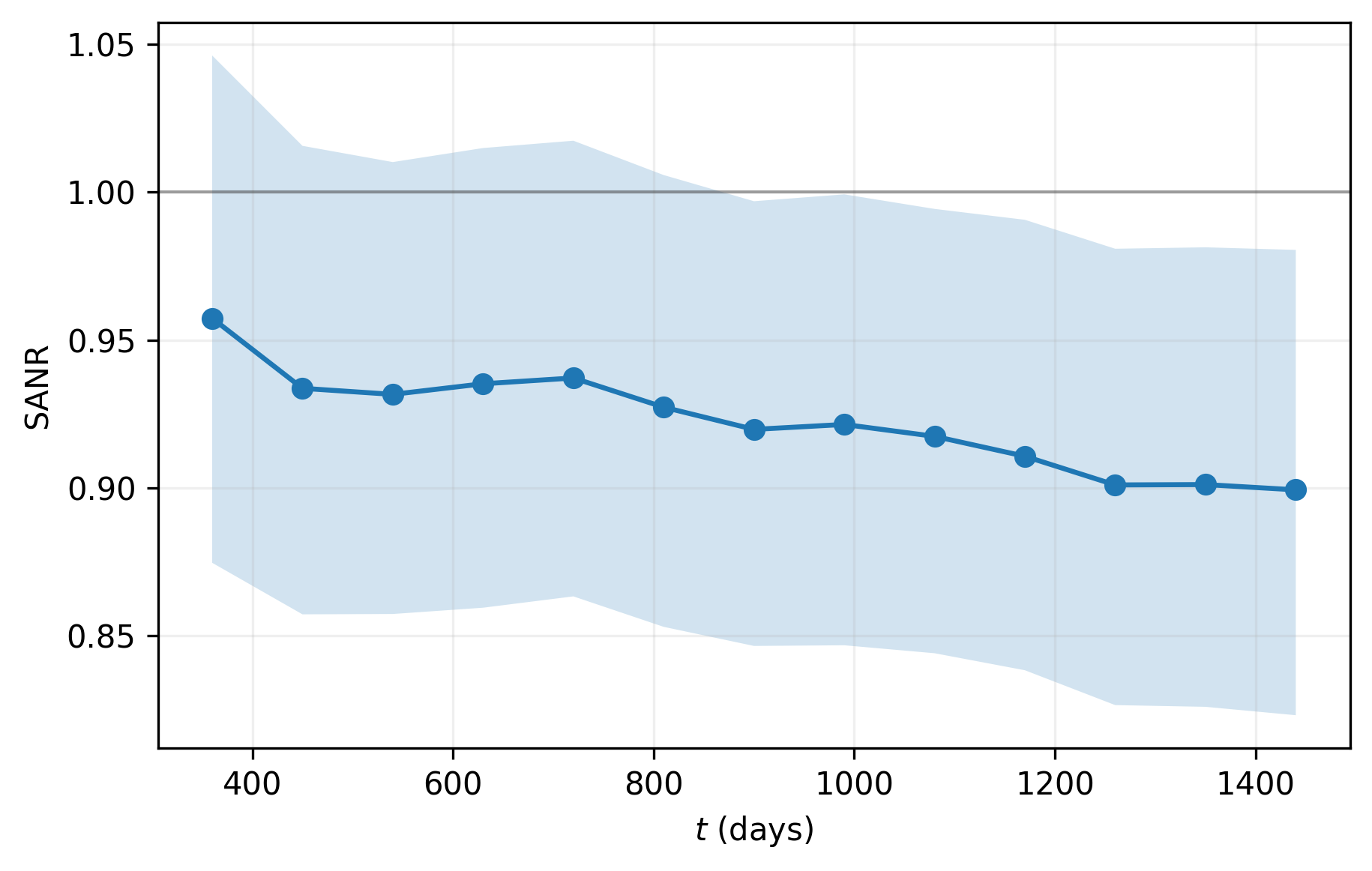}
		\caption{Posterior mean and 95\% posterior interval for $\mathrm{SANR}(t;r)$ with $r=t$, $\rho=0.5$.}
		\label{fig:SANR_diag_HFACT}
	\end{figure}

\begin{figure}
		\centering
		\includegraphics[width=14cm]{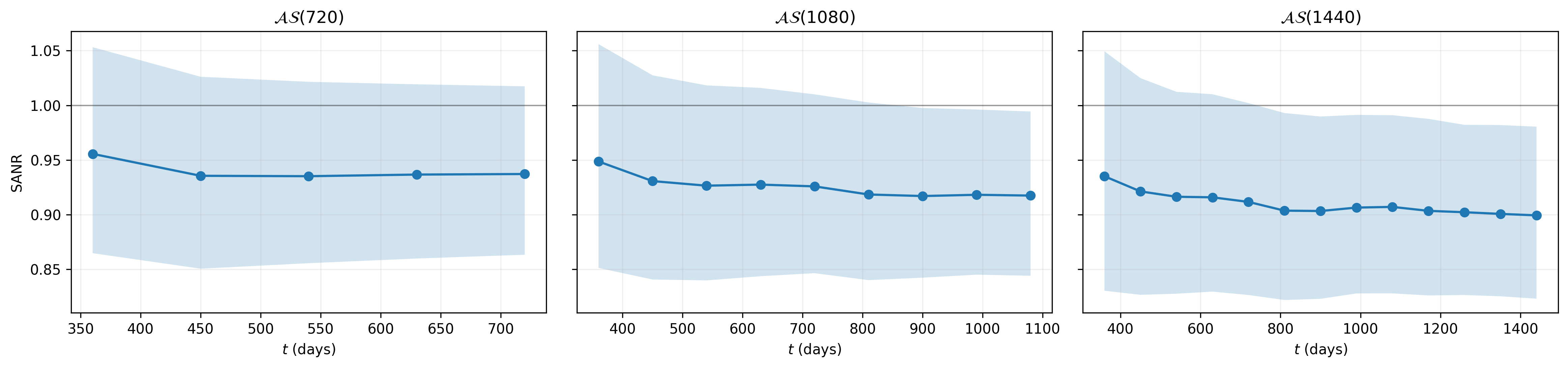}
		\caption{Posterior mean and 95\% posterior interval for $\mathrm{SANR}(t;r)$ as a function of $t$ within fixed always-survivor stratum $\mathcal{AS}(r)$ with $r\in\{720,1080,1440\}$,  $\rho=0.5$.}
		\label{fig:SANR_fixedr_HFACT}
	\end{figure}

	We summarize the posterior distribution of $\mathrm{SANR}(t;r)$ through two informative one-dimensional slices that clarify both the time-evolution of morbidity effects and their dependence on the survivorship index $r$. Figure~\ref{fig:SANR_diag_HFACT} plots $\mathrm{SANR}(t;r)$ along the traditional diagonal slice $r=t$, which corresponds to the conventional ``single-indexed'' always-survivor definition \citep[e.g., ][]{Lyu2023}. Figure~\ref{fig:SANR_fixedr_HFACT} instead fixes $r\in\{720,1080,1440\}$ (i.e., 2, 3, and 4 years) and plots $\mathrm{SANR}(t;r)$ as $t$ increases within the same principal stratum, which is the key interpretive advantage of the proposed double-index formulation.
	Three primary findings emerge. First, posterior means suggest that adding structured exercise leads to a modest but increasing reduction in cumulative hospitalization burden over time. Across the plotted horizons, posterior mean $\mathrm{SANR}(t;r)$ is consistently below 1, indicating fewer expected hospitalizations under structured training among always-survivors. Along the diagonal $r=t$ (Figure~\ref{fig:SANR_diag_HFACT}), the posterior mean declines from 0.957 at $t=360$ days (1 year) to 0.90 by $t = 1440$ days (4 years), corresponding to a 4.3\% reduction at 1 year and a 10.1\% reduction by 4 years in rehospitalization among always-survivors. The fixed-$r$ plots show a similar gradual decline: for example, when $r=1440$, the posterior mean decreases from about 0.935 at $t=360$ to 0.899 by $t=1440$ (Figure~\ref{fig:SANR_fixedr_HFACT}), suggesting a steady, cumulative reduction in the long-term burden of rehospitalization.
	
	Second, under the traditional $r=t$ stratum definition, the cumulative benefit becomes clearly detectable only with sufficiently long follow-up. 
	Although posterior means remain below 1 throughout, uncertainty is substantial in early-to-mid follow-up. As illustrated in Figure~\ref{fig:SANR_diag_HFACT}, the 95\% posterior band includes the null value $\mathrm{SANR}=1$ (equivalently, $\log\mathrm{SANR}=0$) through approximately $t\approx 800$ days, and the upper bound falls below 1 only after roughly $t\gtrsim 900$ days. Thus, under the conventional time-indexed always-survivor definition, the result of HF-ACTION supports the burden-reduction benefit of structured training but only at longer horizons. This pattern is consistent with the broader understanding that the benefits of supervised exercise and other lifestyle interventions often accumulate gradually over time rather than appearing immediately among heart failure patients \citep{Sachdev2023}.
	
	Third, fixing the survivorship stratum reveals meaningful heterogeneity that is obscured under the conventional $r=t$ perspective.
	The fixed-$r$ panels in Figure~\ref{fig:SANR_fixedr_HFACT} show that the strength of evidence can differ across always-survivor strata defined by time limit $r$ even at the same time horizon $t$. In particular, at $t=800$ days the 95\% posterior interval for $\mathrm{SANR}(t;r)$ still includes the null when the stratum is defined at $r=1080$ (middle panel), whereas the corresponding interval lies entirely below 1 when the stratum is defined at $r=1440$ (right panel). In other words, at the same disease burden horizon ($t=800$), restricting attention to the subgroup of patients who would survive to 1440 days under either strategy yields stronger evidence that supervised training reduces rehospitalization than expanding to the broader always-survivor stratum of those who would survive to 800 days under either treatment allocation. This distinction is substantively important for clinical practice: $\mathcal{AS}(1440)$ is a distinct, nested patient subgroup relative to $\mathcal{AS}(1080)$, and the estimated treatment effect on rehospitalization can vary across these latent risk strata. In contrast, this nuanced pattern will not be detected under the traditional $r=t$ definition, because the target subgroup changes with $t$, making it difficult to disentangle time evolution of morbidity effects from changes in the underlying principal stratum.
	
	Overall, traditional principal stratification for censored time-to-event outcomes is often implemented with a single time index, effectively setting the survivorship index equal to the outcome horizon ($r=t$), so that the target subpopulation is $\mathcal{AS}(t)$. In recurrent-event settings, as illustrated by HF-ACTION, an apparent time trend with the single-index convention in $\mathrm{SANR}(t;t)$ can conflate (i) genuine accumulation of treatment differences in recurrent morbidity with (ii) changes in the composition of the always-survivor subgroup as $t$ increases. Our double-index formulation decouples these roles: $r$ fixes a survivorship-defined subgroup $\mathcal{AS}(r)$, while $t$ indexes the cumulative recurrent-event burden within that fixed subgroup. Consequently, plots of $\mathrm{SANR}(t;r)$ as a function of $t$ at fixed $r$ directly describe how disease burden accrues over time for the same latent subgroup, and comparisons across $r$ at a fixed $t$ reveal heterogeneity across the always-survivor stratum under different limits that are not accessible under the traditional $r=t$ definition.

\begin{figure}[tb]
		\centering
		\includegraphics[width=9cm]{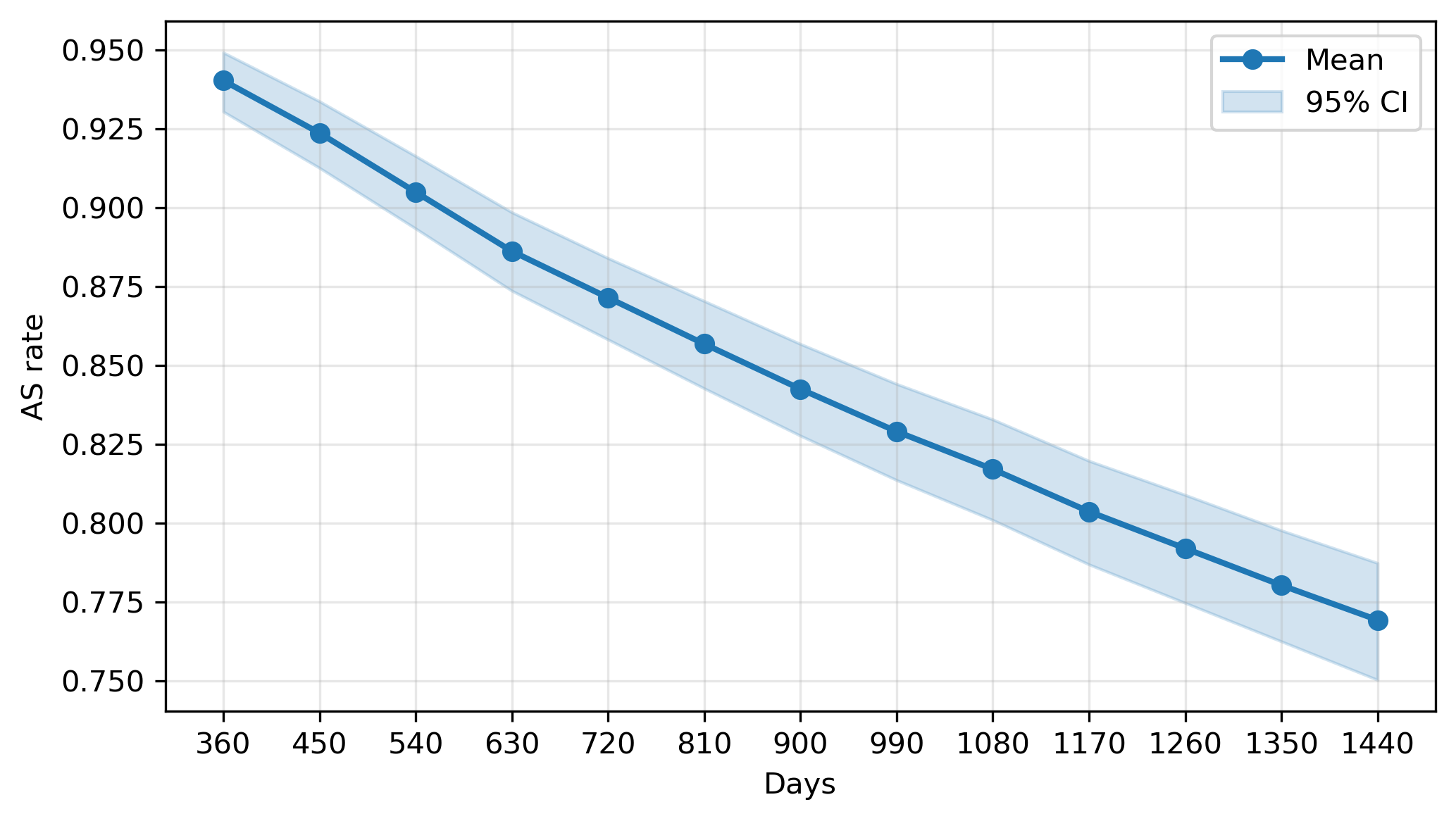}
		\caption{Estimated always-survivor (AS) rate $p(r<D_i^0,r<D_i^1)$ over time (days) in HF-ACTION. Points connected by the solid line indicate the posterior mean AS rate at each time index; the shaded band denotes the corresponding $95\%$ posterior interval.}
		\label{fig:always_survivors}
	\end{figure}

	In addition, Figure~\ref{fig:always_survivors} reports the posterior mean and $95\%$ posterior interval for the always-survivor rate $p(r<D_i^0,r<D_i^1)$ in HF-ACTION, which quantifies the fraction of participants who would survive beyond $r$ under both treatment assignments. The estimated always-survivor rate decreases steadily over time, from approximately $0.94$ at $r=360$ days to approximately $0.77$ at $r=1440$ days, with a modest widening of the posterior interval at longer horizons. Two implications for heart failure care follow. First, even at multi-year horizons the always-survivor stratum remains substantial, so the principal-stratum estimands pertain to a large, clinically relevant portion of the cohort rather than to a diminishing small subgroup. Second, the downward trend reflects the nontrivial mortality in HF-ACTION, and the gradual increase in uncertainty at later time $r$ underscores that cross-world survivorship membership becomes progressively harder to identify as the horizon $r$ grows, motivating explicit sensitivity analysis in the weakly identified cross-world component.
	
	\subsection{Sensitivity analysis for cross-world dependence}\label{sec:SA}
\begin{figure*}[htbp]
		\centering
		\begin{subfigure}[b]{14cm}
			\centering
			\includegraphics[width=4.5cm]{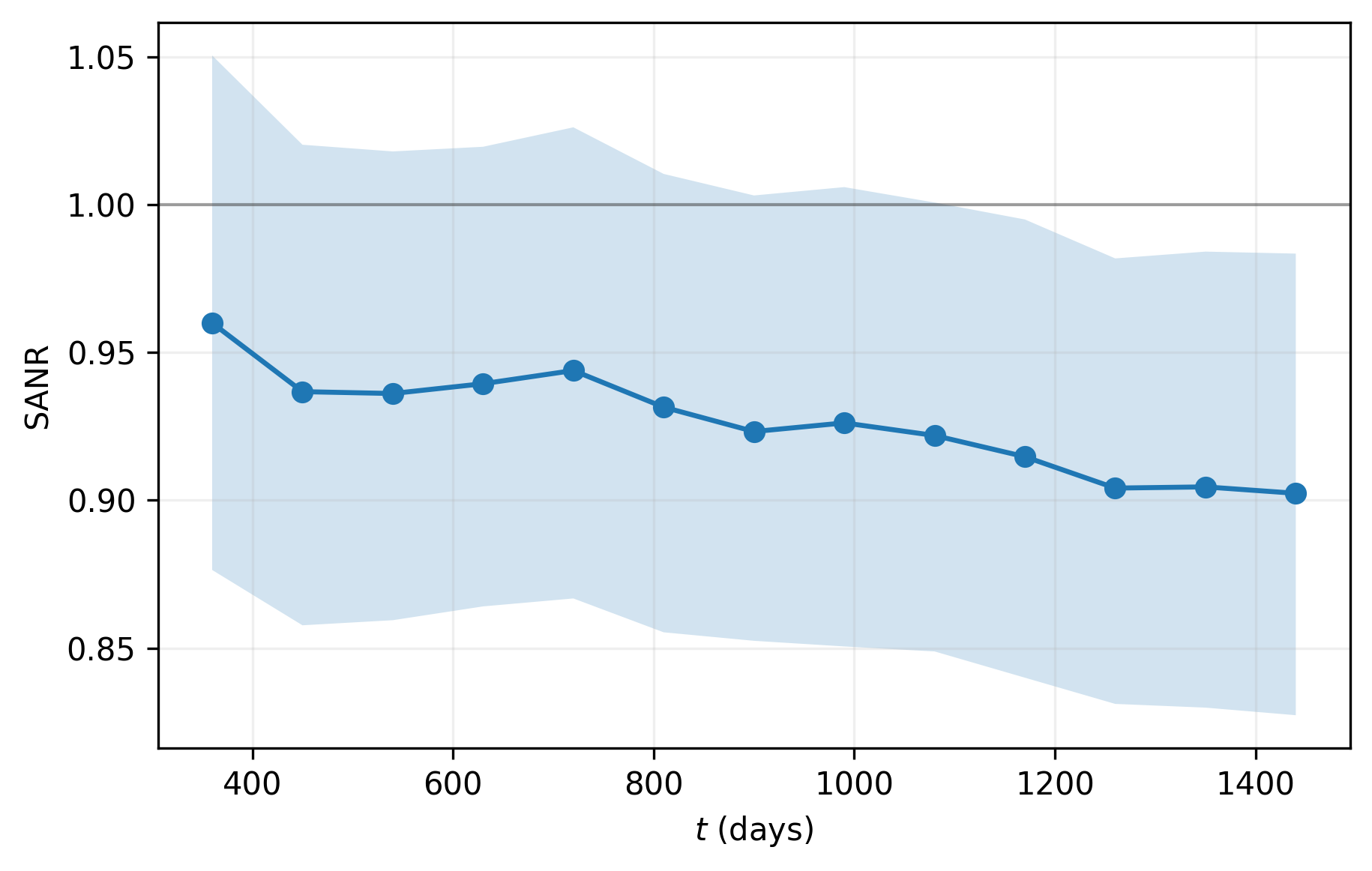}\hfill
			\includegraphics[width=9.cm]{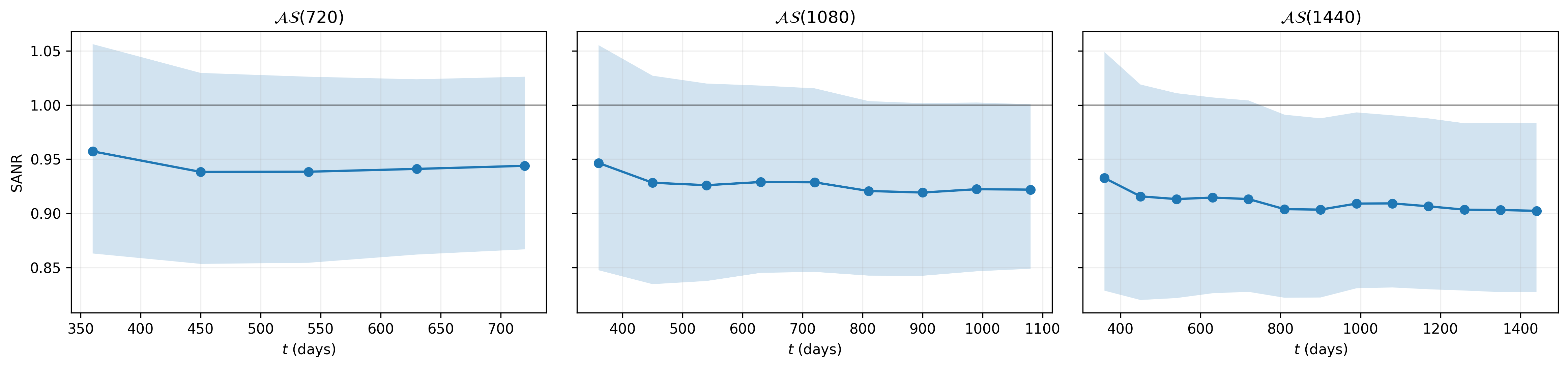}
			\caption{$\rho=0.1$}
		\end{subfigure}
		
		\vspace{0.8em}
		
		\begin{subfigure}[b]{14cm}
			\centering
			\includegraphics[width=4.5cm]{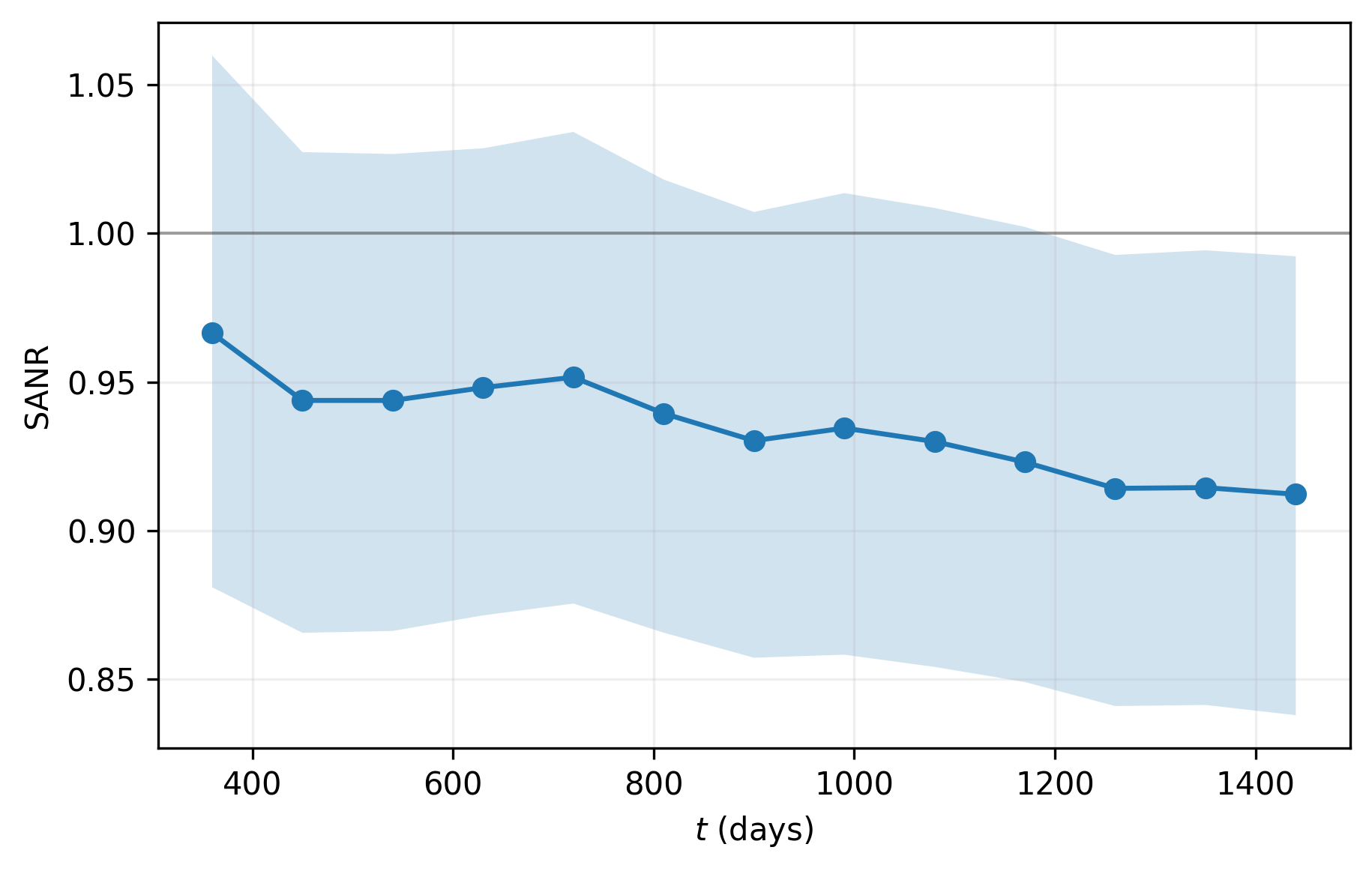}\hfill
			\includegraphics[width=9.cm]{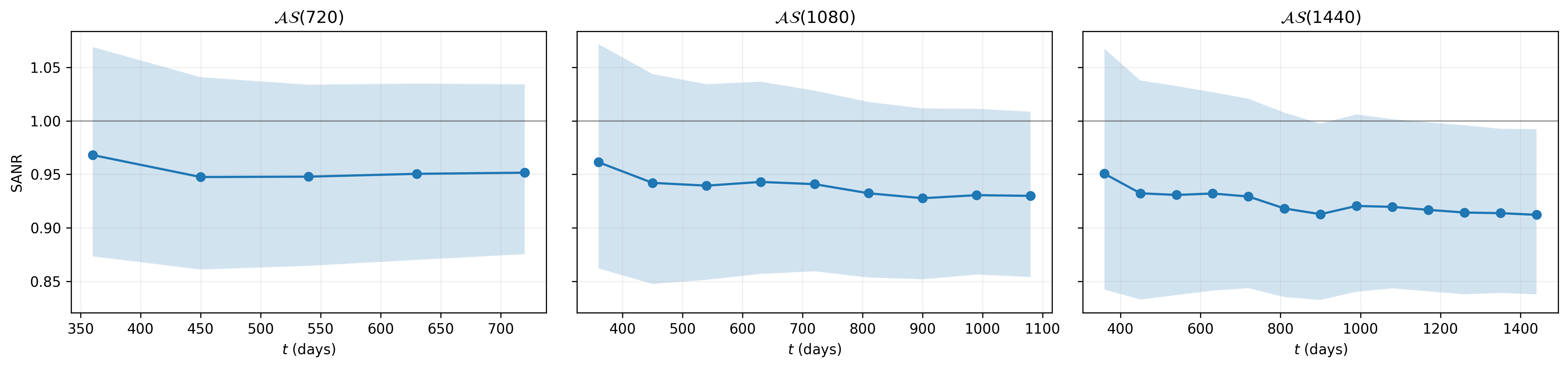}
			\caption{$\rho=0.3$}
		\end{subfigure}
		
		\vspace{0.8em}
		
		\begin{subfigure}[b]{14cm}
			\centering
			\includegraphics[width=4.5cm]{figures/SANR_diag_HFACT_0.5.png}\hfill
			\includegraphics[width=9.cm]{figures/SANR_fixedr_HFACT_0.5.png}
			\caption{$\rho=0.5$}
		\end{subfigure}
		
		\vspace{0.8em}
		
		\begin{subfigure}[b]{14cm}
			\centering
			\includegraphics[width=4.5cm]{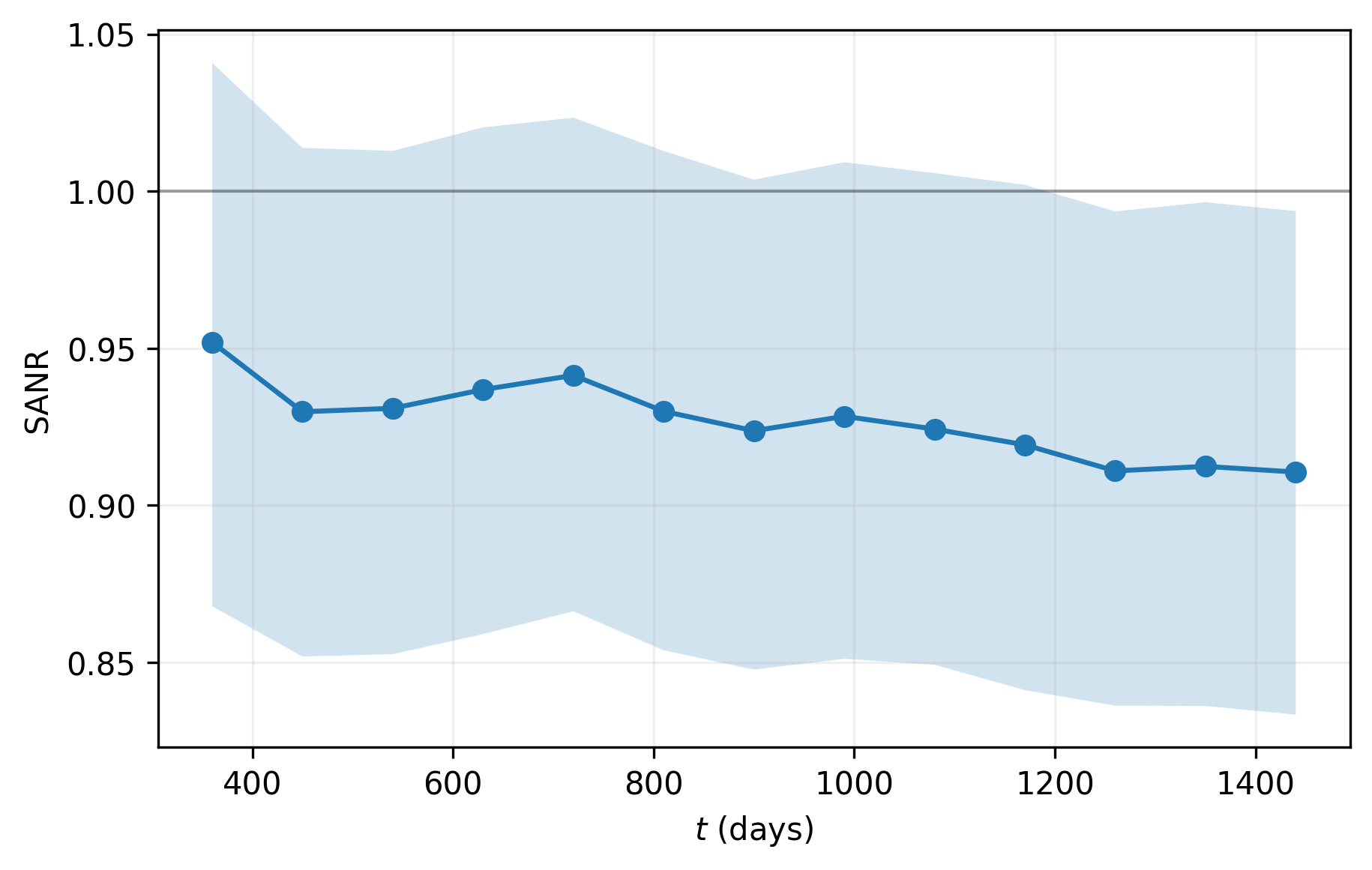}\hfill
			\includegraphics[width=9.cm]{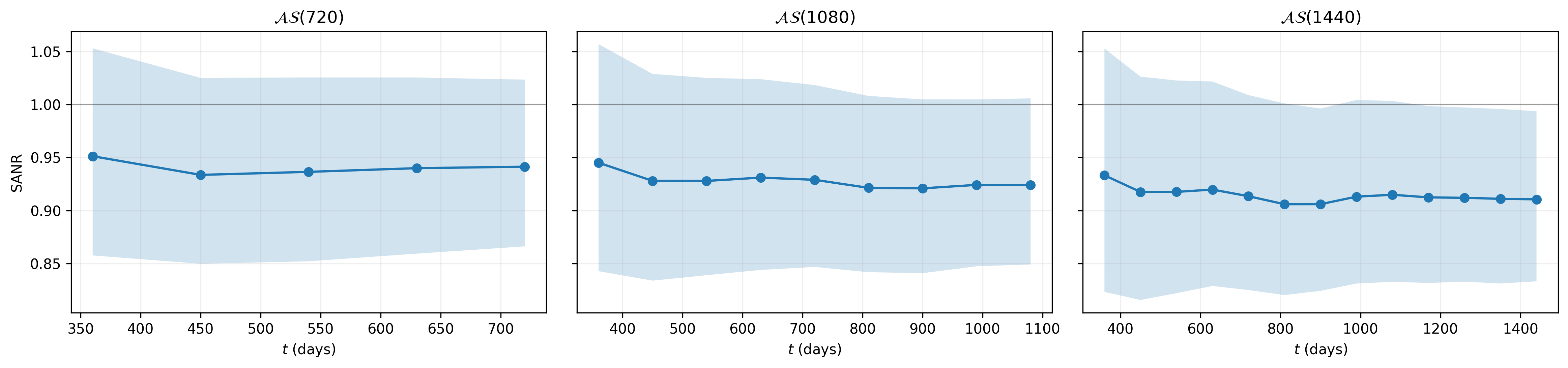}
			\caption{$\rho=0.7$}
		\end{subfigure}
		
		\vspace{0.8em}
		
		\begin{subfigure}[b]{14cm}
			\centering
			\includegraphics[width=4.5cm]{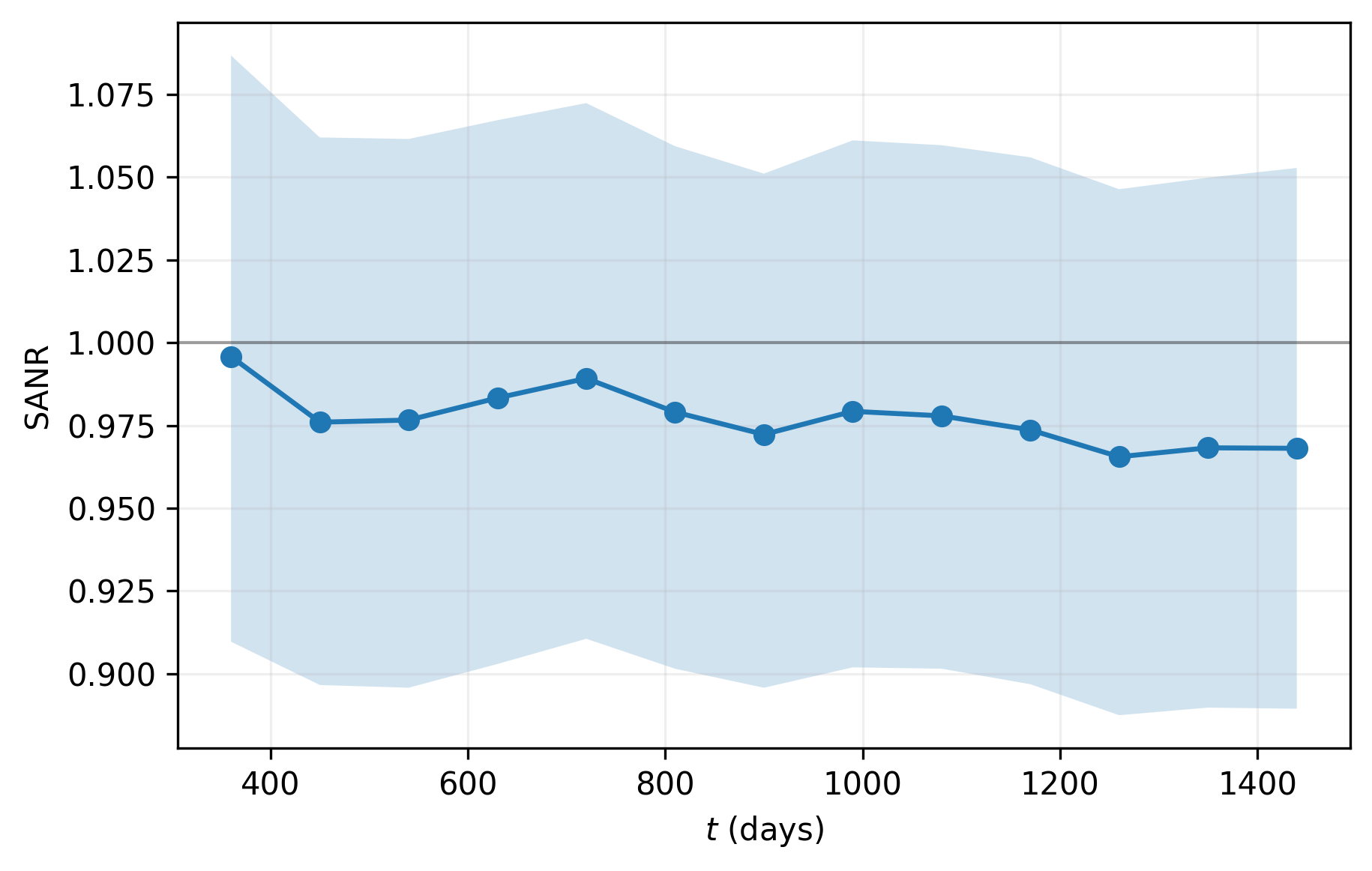}\hfill
			\includegraphics[width=9.cm]{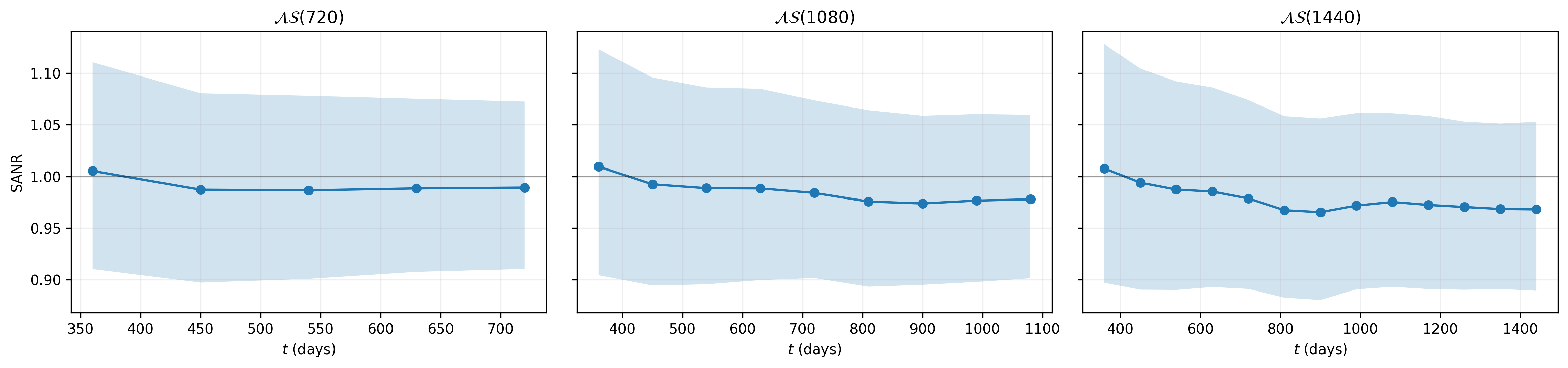}
			\caption{$\rho=0.9$}
		\end{subfigure}
		
		\caption{Posterior mean and $95\%$ posterior interval for $\mathrm{SANR}(t;r)$ in HF-ACTION under five sensitivity parameters $\rho$. In each row, the left panel shows the diagonal slice $r=t$, and the right panel shows fixed always-survivor stratum $r\in\{720,1080,1440\}$.}
		\label{fig:sensitivity_analysis_SANR_trend}
	\end{figure*}

	This section complements the empirical results above by assessing robustness to the cross-world frailty correlation parameter $\rho$, which governs dependence between the two potential processes and is not identified from the observed data. Figures~\ref{fig:sensitivity_analysis_SANR_trend} summarize posterior inference for $\mathrm{SANR}(t;r)$ under five values $\rho\in\{0.1,0.3,0.5,0.7,0.9\}$. We focus on a positive grid because the scientifically plausible concern in this setting is positive cross-world dependence induced by shared latent health status, whereas negative dependence appears much less credible. For each $\rho$, we display both the diagonal slice $r=t$ and fixed-stratum slices with $r\in\{720,1080,1440\}$ in Figure~\ref{fig:sensitivity_analysis_SANR_trend}. 
	For $\rho \in \{0.1,0.3,0.5,0.7\}$, the qualitative pattern is stable: posterior means are below 1 and generally decline with $t$, indicating a consistent direction of effect toward reduced cumulative hospitalization burden under exercise training among always-survivors. Varying $\rho$ between $0.1$ and $0.7$ does not shift the level of the curves and alters posterior uncertainty, thus the substantive interpretation of the HF-ACTION results remains the same.
	For example, under $\rho=0.3$, posterior means of $\mathrm{SANR}(t;r)$ remain below 1 across the plotted horizons and generally decline with $t$, indicating a persistent, modest reduction in cumulative hospitalization burden under exercise training among always-survivors. Along the diagonal slice, the posterior mean decreases from $0.960$ at $t = 360$ days to $0.902$ by $t = 1440$ days. The $95\%$ posterior interval continues to include the null value $\mathrm{SANR}=1$ throughout $t=1080$, which is a slightly longer horizon than the primary analysis under $\rho=0.5$, but the overall trend remains similar.
	The fixed-$r$ summaries also show a similar pattern to the primary analysis.

	However, a somewhat different pattern emerges under the more extreme choice of $\rho=0.9$. A large value of $\rho$ implies very strong cross-world correlation between the frailties $\gamma^0$ and $\gamma^1$, so that the model effectively approximates a single-frailty specification with $\gamma^0 \approx \gamma^1$, similar in spirit to the single shared-frailty formulation adopted in \citet{Lyu2023} and \citet{Comment2025}. Under this more extreme assumption, the $95\%$ posterior interval continues to include the null value $\mathrm{SANR}=1$ throughout the displayed range, with the upper bound remaining close to 1 even at the longest follow-up times. Thus, relative to the primary dual-frailty analysis, imposing near-perfect cross-world dependence yields more conservative inference; although the posterior means remain consistently below 1, the evidence for a reduction in rehospitalization burden becomes less decisive.
	The fixed-$r$ summaries show the same pattern. For each survivorship stratum $r$, the posterior means remain below 1 and decline gradually with $t$, but the $95\%$ posterior intervals typically continue to overlap 1 over the observed follow-up period. In particular, at $t=900$ days, the posterior interval still includes $\mathrm{SANR}=1$ both when the always-survivor stratum is defined at $r=1080$ and when it is defined at $r=1440$. Hence, under $\rho=0.9$, the stratum-specific separation from the null that was visible under the dual-frailty analysis with $\rho=0.1$ to $0.7$ is largely attenuated.
	
	Taken together, these results indicate that the qualitative direction of the estimated exercise effect among always-survivors is robust, in the sense that the posterior mean $\mathrm{SANR}$ remains below 1 across the sensitivity analysis. When $\rho$ lies in the moderate range $\rho=0.1$ to $0.7$, the dual-frailty formulation reveals meaningful stratum-specific evidence of reduced rehospitalization burden. By contrast, when $\rho=0.9$, the near single-frailty assumption suppresses that signal and pulls the inference toward the null. This attenuation is substantively important; in our setting, treatment may affect rehospitalization risk and mortality risk through partially distinct latent pathways, so forcing the two cross-world frailties to be nearly identical runs the risk of masking clinically meaningful heterogeneity. For this reason, the dual-frailty framework is not only a flexible generalization of prior shared-frailty approaches, but also an important modeling device for recovering treatment-effect patterns that would otherwise be obscured under an overly restrictive dependence assumption.

	\section{Concluding Remarks}
	\label{sec:conclusion}
	
	Motivated by HF-ACTION, a real-world randomized clinical trial where the endpoint of the number of rehospitalizations is clinically important but subject to truncation by death, we proposed a new principal stratification approach for causal recurrent event analysis. Our work makes two main contributions. First, we introduce a new principal causal estimand for recurrent events,together with a partial identification result that enables transparent sensitivity analysis for the cross-world dependence structure of potential outcomes. Second, we develop the EDDPM prior, a nested Bayesian nonparametric construction that jointly models recurrent and terminal events while embedding covariate dependence and naturally accommodating the frailty structure required for identification. By separating terminal-event
	clusters from finer recurrent-event subclusters, the EDDPM addresses the unfavorable random partition behavior of standard DP priors, and offers flexibility over parametric approaches \citep{Nevo2022,Lyu2023}. We provided a fully tractable Gibbs sampling algorithm for posterior inference and demonstrated through simulation studies that the proposed method outperforms traditional DP-based and parametric approaches in both accuracy and robustness. 
	
	Beyond these methodological contributions, the HF-ACTION analysis also illustrates why the proposed framework is useful compared to existing alternatives in practice. Above all, the double-index estimand $\mathrm{SANR}(t;r)$ yields different and richer conclusions than the conventional single-index alternative. Because enforcing $r = t$ makes the always-survivor subgroup shrink as the outcome horizon advances, any apparent time trend inevitably mixes genuine changes in rehospitalization dynamics with compositional shifts in the at-risk population, whereas decoupling
	$r$ from $t$ holds the target subpopulation fixed and reveals useful heterogeneity across always-survivor strata
	that is entirely invisible under the traditional definition. Second, our sensitivity analysis approach for the cross-world frailty correlation $\rho$ shows that the choice between a dual-frailty and a single-frailty specification is not merely a matter of modeling generality but can change the substantive conclusions drawn from the data. That is, at moderate values of $\rho$ the dual-frailty formulation detects a gradually accumulating reduction in rehospitalization burden that eventually separates from the null, whereas forcing the two frailties to be nearly identical (as assumed in existing shared-frailty approaches in \citet{Lyu2023}) attenuates
	this signal. Taken together, these two features of the analysis highlight the applied value of the proposed framework for chronic-disease trials in which recurrent morbidity and mortality are intertwined. Our HF-ACTION case study encourages practitioners to move beyond single-index, single shared-frailty analyses when the scientific question concerns cumulative disease burden among always-survivors.

	\section*{Acknowledgement}
	Research in this article was supported by the United States National Institutes of Health (NIH), National Heart, Lung, and Blood Institute (NHLBI, grant number R01-HL168202 and 1R01HL178513).

	\bibliographystyle{Chicago}
	\bibliography{literature}

	\newpage
	\appendix

	\section{A potential pitfall of DP}
	\label{sec:potential_pitfall_dp}
	\subsection{DP for the recurrent event analysis with a terminal event}
	As recent non-causal descriptive analyses for recurrent processes with a terminal event, \citet{Paulon2020} jointly modeled the gap times of recurrent and terminal events using a Dirichlet process mixture model with log-normal kernels for both the subject-specific frailty parameter and the error term.
	\citet{Tian2024} used a DP prior for the scale parameter of Gumbel (extreme value distribution) error terms for the log-scaled survival time.  
	These joint DP models of the log-transformed gap times and survival times, i.e., $ Y^z_{ij} = \log(W^z_{ij}) $ and $ U^z_i = \log(D^z_i) $, are typically specified as follows:
	\begin{equation}
		\label{eq:dmp_joint}
		U_i, \overline{Y}_i \mid P \sim f\qty(u, \overline{y} \mid P) = \int K(u, \overline{y} \mid \xi) dP(\xi) =\sum_{j=1}^{\infty} w_j K(u, \overline{y} \mid \xi_j),
	\end{equation}
	where $\overline{Y}_i = (Y_{i1},\ldots, Y_{iN_i})^\top$, $U_i$ is univariate, $K(\cdot)$ is an appropriate kernel function, and the mixing distribution $P$ is given a DP prior with scale parameter $\alpha>0$ and base measure $P_0$, denoted by $P \sim \mathrm{DP}(\alpha, P_0)$. The second equation follows from the a.s. discrete nature of the DP, that is, the model reduces to a countable mixture with the mixing weights $w_j$ having a stick breaking prior with parameter $\alpha$ and $\xi_j \sim P_0$ independently of the $w_j$.
	This model can be alternatively described as the following hierarchical model: 
	$ U_i \mid \theta_i \sim F_u(\cdot \mid \phi_i, \gamma_i)$, $\overline{Y}_i \mid \theta_i \sim F_{\overline{y}}(\cdot \mid \theta_i, \gamma_i)$, $(\phi_i, \theta_i, \gamma_i) \mid P \sim P$, $ P \sim \mathrm{DP}(\alpha, P_{0\phi} \times P_{0\theta} \times P_{0\gamma} ),$
	where the atom $\xi_i$ is decomposed into $\phi_i$ and $\theta_i$, which are survival and recurrent parameters, respectively. Here the base measure $ P_{0\phi} $ and $ P_{0\theta} $ of the DP assumes independence between $\phi$ and $\theta$.  We further assume that $ P_{0\phi} $ and $ P_{0\theta} $ are  absolutely continuous, admitting densities $ p_{0\phi} $ and $ p_{0\theta} $. Then \eqref{eq:dmp_joint} can be written as $f\qty(u, \overline{y} \mid P) = \sum_{j=1}^{\infty} w_j K(u \mid \overline{y}, \phi_j)K( \overline{y} \mid \theta_j)$,
	where $K(u \mid \overline{y}, \phi)$ and $K( \overline{y} \mid \theta)$ are the densities associated to $F_u( \cdot \mid \overline{y}, \phi)$ and $F_{\overline{y}}( \cdot \mid \theta)$.
	
	While this is a simple, common approach adopted in literature for the joint analysis of recurrent events with a terminal event \citep{Juhee2019, Paulon2020, Xu2021, Willem2022, Xu2022, Tian2024, Zehavi2025}, a potential pitfall of these joint specifications is that the latent random partition induced by the DP will be overwhelmingly determined by the recurrent events rather than the terminal event as the number of recurrences grows. This issue occurs because the likelihood contribution of the recurrent event dominates that of a single terminal event, which may not even be observed due to other censoring events. Consequently, the DP prior favors a random partition that approximates the distribution of the recurrent events well with as many clusters as necessary. However, having too many small clusters, which are dominantly determined by the contribution of $\overline{y}$, can lead to unreliable within-cluster predictions of the survival event $u$, with higher posterior variance due to unnecessarily small sample sizes, thereby compromising the predictive performance for the survival event. Predictive performance is crucial for Bayesian causal inference because the inferential procedure imputes missing potential outcomes and, therefore, heavily depends on the accuracy of potential outcome predictions.
	Additionally, existing approaches use the joint DP primarily to model a frailty term. However, because the frailty is widely recognized as a subject-specific distinct parameter that reflects an individual's vulnerability (e.g., physical health conditions), it is somewhat counterintuitive that the random partition, which yields the subject-specific frailty, is driven largely by numerous recurrent events rather than the subject-specific survival event.

	\subsection{Random partition induced by the standard DP}
	\label{sec:random_partition_dp}
	This section discusses the adverse impact of the standard DP random partition on predictive performance for recurrent outcomes with truncation by death.
	Let $\mathcal{P}_n = (s_1, \dots, s_n)$ be the random partition defined by cluster allocation labels, where $s_i = j$ if $(\phi_i, \theta_i)$ equals the $j$-th unique value $(\phi_j^*, \theta_j^*)$, for $j = 1, \ldots, M_n$. 
	Here, $M_n$ is the number of distinct groups (clusters) in the partition $\mathcal{P}_n$. For each cluster $j$, let $\mathcal{S}_j = \{ i : s_i = j \}$
	be the set of indices for individuals assigned to the $j$-th cluster. Define $\overline{Y}_j^* = \{\overline{Y}_i : i \in \mathcal{S}_j\}$, $U_j^* = \{U_i : i \in \mathcal{S}_j\}$, and let $\overline{Y}_{1:n} = (\overline{Y}_1, \dots, \overline{Y}_n)$, $U_{1:n} = (U_1, \dots, U_n)$.
	
	Under this setup, the joint posterior distribution of the random partition $\mathcal{P}_n$ and the cluster-specific parameters  $(\phi^*, \theta^*) = (\phi_j^*, \theta_j^*)_{j=1}^{M_n}$ is given by
	\begin{equation}
		\label{eq:joint_post_part_parameters}
		p(\mathcal{P}_n, \phi^*, \theta^* \mid \overline{Y}_{1:n}, U_{1:n})
		\propto 
		p(\mathcal{P}_n)
		\prod_{j=1}^{M_n} p_{0\phi}(\phi_j^*)  p_{0\theta}(\theta_j^*) 
		\prod_{j=1}^{M_n} \prod_{i \in \mathcal{S}_j} K\left(U_i \mid \overline{Y}_i, \phi_j^*\right)
		K\left(\overline{Y}_i \mid \theta_j^*\right).
	\end{equation}
	The prior on the random partition induced by the DP is $p(\mathcal{P}_n) \propto \alpha^{M_n} 
	\prod_{j=1}^{M_n} \Gamma(n_j),$
	where $n_j = |\mathcal{S}_j|$ \citep{Antoniak1974} and $\alpha$ is a precision parameter of the DP. Consequently, the posterior distribution of the random partition $\mathcal{P}_n$ can be written as
	\begin{equation}
		\label{eq:post_random_partition}
		p(\mathcal{P}_n \mid \overline{Y}_{1:n}, U_{1:n})
		\propto
		\alpha^{M_n} \prod_{j=1}^{M_n} \Gamma(n_j) 
		h_u\left(U_j^* \mid \overline{Y}_j^*\right)
		h_y\left(\overline{Y}_j^*\right),
	\end{equation}
	where $h_u\left(U_j^* \mid \overline{Y}_j^*\right) 
	= \int_{\Phi} \prod_{i \in \mathcal{S}_j} K\left(U_i \mid \overline{Y}_i, \phi\right)
	dP_{0\phi}(\phi)$ and $h_y\left(\overline{Y}_j^*\right) 
	= \int_{\Theta} \prod_{i \in \mathcal{S}_j} K\left(\overline{Y}_i \mid \theta\right)
	dP_{0\theta}(\theta)$.
	From \eqref{eq:joint_post_part_parameters}, the posterior distributions of the cluster-specific parameters are
	\begin{equation}
		\label{eq:post_parameters}
		\begin{split}
			p\left(\phi_j^* \mid \mathcal{P}_n, \overline{Y}_{1:n}, U_{1:n}\right)
			&\propto 
			p_{0\phi}\left(\phi_j^*\right)
			\prod_{i \in \mathcal{S}_j} 
			K\left(U_i \mid \overline{Y}_i, \phi_j^*\right),\\[6pt]
			p\left(\theta_j^* \mid \mathcal{P}_n, \overline{Y}_{1:n}, U_{1:n}\right)
			&\propto 
			p_{0\theta}\left(\theta_j^*\right)
			\prod_{i \in \mathcal{S}_j} 
			K\left(\overline{Y}_i \mid \theta_j^*\right).
		\end{split}
	\end{equation}
	\noindent
	\eqref{eq:post_random_partition} indicates that, under the DP partition, individuals are grouped according to similarities in recurrent outcomes and their relationship to the survival outcome. 
	When multiple recurrences are observed, the likelihood term involving $\overline{Y}$ often dominates the posterior in 
	\eqref{eq:post_random_partition}. 
	However, from \eqref{eq:post_parameters}, the posteriors of each cluster's parameters $\phi_j^*$ and $\theta_j^*$ are updated by observations of all individuals in the shared subset $\mathcal{S}_j$. 
	Consequently, the posterior of $\phi_j^*$ may fail to capture local trends of survival events if the partition is driven largely by the recurrent events. 
	Moreover, if many clusters are needed to approximate the distribution of recurrent outcomes (i.e., if $M_n$ is large), 
	each cluster may contain relatively few observations. 
	In that case, the posterior for $\phi_j^*$ is inferred from an unnecessarily small sample, yielding an unreliable mean estimate and large posterior variance, with the prior continuing to exert strong influence.

	\section{Proofs}
	\label{sec:proofs}
	\subsection{Proof of Proposition \ref{prop:1}}
	\begin{proof}
		Independence of conditional distributions on $\phi \in \Phi$ implies that
		\begin{equation}
			\label{eq:proof1}
			p(\mathcal{P}_n, \phi^*) = p(\mathcal{P}_{n,u}) \prod_{j=1}^{M_n} p_{0\phi}\left(\phi_j^*\right)p(\mathcal{P}_{n,y} \mid \mathcal{P}_{n,u}, \phi_j^*) = p(\mathcal{P}_{n,u}) \prod_{j=1}^{M_n} p_{0\phi}\left(\phi_j^*\right)p(\mathcal{P}_{n,y} \mid \phi_j^*)
		\end{equation}
		The result of \citet{Antoniak1974} implies that the random partition induced by the DP is $p(\mathcal{P}_{n,u}) = \frac{\Gamma(\alpha_{\phi})}{\Gamma(\alpha_{\phi}+n)} \alpha_{\phi}^{M_n} \prod_{j=1}^{M_n}\Gamma(n_j) $ 
		and 
		$p(\mathcal{P}_{n,y} \mid \phi_j^*) = \frac{\Gamma(\alpha_{\theta}(\phi_j^*))}{\Gamma(\alpha_{\theta}(\phi_j^*) + n_j)} \alpha_{\theta}(\phi_j^*)^{M_{n,j}} \prod_{l=1}^{M_{n,j}}\Gamma(n_{l \mid j})$. Putting these into \eqref{eq:proof1} and integrating out $\phi^*$, we obtain that the prior on the random partition induced by the EDP is 
		\begin{align*}
			p(\mathcal{P}_n) = \frac{\Gamma(\alpha_{\phi})}{\Gamma(\alpha_{\phi}+n)} \alpha_{\phi}^{M_n} \prod_{j=1}^{M_n} \int_{\Phi} \alpha_{\theta}(\phi)^{M_{n,j}} \frac{\Gamma(\alpha_{\theta}(\phi)) \Gamma(n_j)}{\Gamma(\alpha_{\theta}(\phi) + n_j)} dP_{0\phi}(\phi) \prod_{l=1}^{M_{n,j}}\Gamma(n_{l \mid j}).
		\end{align*}
		The claim in the proposition can be obtained by the Bayes theorem.
	\end{proof}

	\subsection{Proof of Theorem \ref{thm:identification_mu}}
	\begin{proof}
		We consider the identification of \eqref{eq:estimand_1}.
		For any $r \in (t, C^{*}]$, we have
		\begin{align*}
			\pr \left( N_i^z(t) = n \mid \mathcal{AS}(r) \right)  
			&= \pr \left( N_i^z(t) = n \mid r < D_i^1, r < D_i^0 \right)  \\
			&= \E_{X,\boldsymbol{\gamma} \mid \mathcal{AS}(r)}\left[\pr\left(N_i^z(t) = n \mid r < D_i^1, r < D_i^0, X, \boldsymbol{\gamma} \right) \right] \\ 
			&= \E_{X,\boldsymbol{\gamma} \mid \mathcal{AS}(r)}\qty[\frac{\pr\qty(N_i^z(t) = n, r < D_i^1, r < D_i^0 \mid X,\boldsymbol{\gamma} )}{\pr\qty(r < D_i^1, r < D_i^0 \mid X,\boldsymbol{\gamma} )} ] \\
			&= \E_{X,\boldsymbol{\gamma} \mid \mathcal{AS}(r)}\qty[\frac{\pr\qty(N_i^z(t) = n, r < D_i^z \mid X,\gamma^{z} ) \pr\qty(r < D_i^{1-z} \mid X,\gamma^{1-z} )}{\pr\qty(r < D_i^1 \mid X,\gamma^1 ) \pr\qty(r < D_i^0 \mid X,\gamma^0 )} ] \\
			&= \E_{X,\boldsymbol{\gamma} \mid \mathcal{AS}(r)}\qty[ \pr\qty(N_i^z(t) = n \mid  r < D_i^z, X,\gamma^{z} )  ] \\
			&= \E_{X,\boldsymbol{\gamma} \mid \mathcal{AS}(r)}\qty[ \pr\qty(N_i^z(t) = n \mid Z_i=z, r < D_i^z, X,\gamma^z )  ] \\
			&= \E_{X,\boldsymbol{\gamma} \mid \mathcal{AS}(r)}\qty[ \pr\qty(N_i(t) = n \mid Z_i=z, r < D_i, X,\gamma^z )  ] \\
			&= \int_{\mathcal{X}}\int_{\Gamma} \pr\qty(N_i(t) = n \mid Z_i=z, r < D_i, X,\gamma^z ) f_{X,\boldsymbol{\gamma} \mid \mathcal{AS}(r)}(x,\boldsymbol{\gamma})d\boldsymbol{\gamma} dx.
		\end{align*} 
		The second line is by the law of iterated expectations, the fourth line is by Assumption \ref{asmp:indep_given_fraity}, the sixth line is by Assumption \ref{asmp:ignorability}, and the seventh line is by Assumption \ref{asmp:consistency}. Now note that
		\begin{align*}
			& f_{X,\boldsymbol{\gamma} \mid \mathcal{AS}(r)}(x,\boldsymbol{\gamma}) = \frac{\pr\qty(r < D_i^1, r < D_i^0 \mid \mathbf{X}_i=\mathbf{x}, \boldsymbol{\gamma} ) f_{\boldsymbol{\gamma}}(\boldsymbol{\gamma})f_{X}(x)}{\int_{\mathcal{X}}\int_{\Gamma} \pr\qty(r < D_i^1, r < D_i^0 \mid \mathbf{X}_i=\mathbf{x}', \boldsymbol{\gamma}' ) f_{\boldsymbol{\gamma}}(\boldsymbol{\gamma}')f_{X}(x') d\boldsymbol{\gamma}'dx'} \\
			&= \frac{\pr\qty(r < D_i^1 \mid \mathbf{X}_i=\mathbf{x}      , \gamma^1 ) \pr\qty( r < D_i^0 \mid \mathbf{X}_i=\mathbf{x}, \gamma^0 ) f_{\boldsymbol{\gamma}}(\boldsymbol{\gamma})f_{X}(x)}{\int_{\mathcal{X}}\int_{\Gamma} \pr\qty(r < D_i^1 \mid \mathbf{X}_i=\mathbf{x}', \gamma^{1'} )\pr\qty( r < D_i^0 \mid \mathbf{X}_i=\mathbf{x}', \gamma^{0'} ) f_{\boldsymbol{\gamma}}(\boldsymbol{\gamma}')f_{X}(x') d\boldsymbol{\gamma}'dx'} \\
			&= \frac{\pr\qty(r < D_i^1 \mid Z_i=1,\mathbf{X}_i=\mathbf{x}, \gamma^1 ) \pr\qty( r < D_i^0 \mid Z_i=0,\mathbf{X}_i=\mathbf{x}, \gamma^0 ) f_{\boldsymbol{\gamma}}(\boldsymbol{\gamma})f_{X}(x)}{\int_{\mathcal{X}}\int_{\Gamma} \pr\qty(r < D_i^1 \mid Z_i=1,\mathbf{X}_i=\mathbf{x}', \gamma^{1'} )\pr\qty( r < D_i^0 \mid Z_i=0,\mathbf{X}_i=\mathbf{x}', \gamma^{0'} ) f_{\boldsymbol{\gamma}}(\boldsymbol{\gamma}')f_{X}(x') d\boldsymbol{\gamma}'dx'} \\
			&= \frac{\pr\qty(r < D_i \mid Z_i=1,\mathbf{X}_i=\mathbf{x}  , \gamma^1 ) \pr\qty( r < D_i \mid Z_i=0,\mathbf{X}_i=\mathbf{x}, \gamma^0 ) f_{\boldsymbol{\gamma}}(\boldsymbol{\gamma})f_{X}(x)}{\int_{\mathcal{X}}\int_{\Gamma} \pr\qty(r < D_i \mid Z_i=1,\mathbf{X}_i=\mathbf{x}', \gamma^{1'} )\pr\qty( r < D_i \mid Z_i=0,\mathbf{X}_i=\mathbf{x}', \gamma^{0'} ) f_{\boldsymbol{\gamma}}(\boldsymbol{\gamma}')f_{X}(x') d\boldsymbol{\gamma}'dx'}\\
			&= \frac{\eta_{r}(1, x, \gamma^1) \eta_{r}(0, x, \gamma^0)  f_{\boldsymbol{\gamma}}(\boldsymbol{\gamma})f_{X}(x)}{\int_{\mathcal{X}}\int_{\Gamma} \eta_{r}(1, x', \gamma^{1'}) \eta_{r}(0, x', \gamma^{0'})  f_{\boldsymbol{\gamma}}(\boldsymbol{\gamma}')f_{X}(x') d\boldsymbol{\gamma}'dx'},
		\end{align*}
		where the first line follows from Bayes' theorem, the second line follows from Assumption \ref{asmp:indep_given_fraity}, the third line follows from Assumption \ref{asmp:ignorability}, and the fourth line follows from Assumption \ref{asmp:consistency}. 
	\end{proof}
	
	\section{Details of Gibbs sampler}
	\label{sec:gibbs_details}
	The posterior distributions of the parameters are obtained from the Markov chain Monte Carlo method. We develop a fully tractable Gibbs sampler that uses the data augmentation method to impute truncated survival and gap times for each unit at both treatment arms and the cluster memberships induced by the EDDP, and exploit the complete likelihood to update the parameters.  Specifically, we iterate between drawing from the conditional distributions of model parameters, potential outcomes, and latent nested cluster memberships given the other variables, respectively. The essential algorithm proceeds as follows:
	\begin{enumerate}[noitemsep]
		\item Given all model parameters, $G_i$ and $H_i$, sample $Y^{*}_{i(N_i+1)}$ and $U^{*}_i$.
		\item Given all model parameters, $Y^{*}_{i(N_i+1)}$ and $U^{*}_i$, sample $G_i$ and $H_i$.
		\item Given $Y^{*}_{i(N_i+1)}$, $U^{*}_i$, $G_i$ and $H_i$, sample all model parameters.
		\item Compute the estimands.
	\end{enumerate}
	For simplicity, in what follows,  we denote by $\mathbf{X}_i$ the augmented covariates that contain the baseline covariates and treatment variable $Z_i$ and denote by  $\mathbf{X}_{ij}$ the covariates that additionally include the time-indexed covariates at the $j$-th recurrent period $\mathbf{V}_{ij}$.

	\subsection{Imputation of $U$}
	\noindent
	For each subject $i$ with $\delta_i^C = 1$, given $G_i$, $\boldsymbol{\gamma}=(\gamma^{Z_i}_1,\ldots,\gamma^{Z_i}_K)$,  $\boldsymbol{\beta_{u}}=(\beta_{u,1},\ldots,\beta_{u,K})$, $\boldsymbol{\tau}^2=(\tau_{1}^2,\ldots,\tau_{K}^2)$ and observed variables, we sample
	\begin{align*}
		U_i \mid - \sim 
		\mathrm{TN}\qty(
		\mathbf{X}_i^\top \beta_{u,G_i} + \gamma^{Z_i}_{G_i}, \tau_{G_i}^2, U_{i}^{\mathrm{obs}}, \infty),
	\end{align*}
	where $\mathrm{TN}(\mu,\sigma^2,l,u)$ denotes the truncated normal distribution with the mean, variance, lower bound, and upper bound parameters.
	\noindent
	If $\delta_i^C=0$ (death), $U_i$ remains unchanged. 
	
	\subsection{Imputation of $Y_{i(N_i+1)}$}
	\noindent
	For each subject $i=1,\ldots,n$, given $G_i$, $H_i$, 
	$\boldsymbol{\gamma}$,  
	$\boldsymbol{\beta_{y}}=\qty((\beta_{y,1 \mid 1},\ldots,\beta_{y,1 \mid K})^\top, \ldots,  (\beta_{y,L \mid 1},\ldots,\beta_{y,L \mid K})^\top)$, 
	$\boldsymbol{\sigma}^2= \qty((\sigma_{1 \mid 1}^2,\ldots,\sigma_{1 \mid K}^2)^\top, \ldots, (\sigma_{L \mid 1}^2,\ldots,\sigma_{L \mid K}^2)^\top)$,
	$\boldsymbol{\psi}= \qty((\psi_{1 \mid 1},\ldots,\psi_{1 \mid K})^\top, \ldots, (\psi_{L \mid 1},\ldots,\psi_{L \mid K})^\top)$,
	and observed variables, we sample
	\begin{align*}
		Y_{i(N_i+1)} \mid - \sim 
		\mathrm{TN}\qty(
		\mathbf{X}_{i(N_i+1)}^\top \beta_{y,H_i \mid G_i}
		+\psi_{H_i \mid G_i}\gamma^{Z_i}_{G_i},
		\sigma_{H_i \mid G_i}^2,Y_{i(N_i+1)}^{\mathrm{obs}},\infty),
	\end{align*}
	where $Y_{i(N_i+1)}^{\mathrm{obs}} = \log(\mathcal{T}_i-T_{iN_i})$.
	
	\subsection{Update of $\mathbf{G}$ and $\mathbf{H}$}
	\noindent
	For each $i$, given all parameters and observed/imputed $U_i$ and $Y_{ij}$,
	\begin{align*}
		p\qty(G_i = k \mid -)
		\propto
		w_k^{\phi} \times
		\underbrace{\mathrm{N}\qty(U_i \mid 
			\mathbf{X}_i\beta_{u,k} + \gamma^{Z_i}_k,\tau_k^2)}_{\text{survival part}}
		\times
		\prod_{j=1}^{N_i+1}
		\underbrace{\sum_{l=1}^{L}
			\qty[w_{l \mid k}^{\theta} \mathrm{N}\qty(
			Y_{ij}\mid \mathbf{X}_{ij}\beta_{y,l \mid k}+\psi_{l \mid k}\gamma^{Z_i}_k,\sigma_{l \mid k}^2)
			]}_{\text{recurrent part}}.
	\end{align*}
	Normalize over $k=1,\dots,K$ to obtain a categorical distribution and sample $G_i$. Then, for each $i$ and $j$, we sample
	\begin{align*}
		p\qty(H_{ij} = l \mid -)
		\propto
		w_{l \mid G_i}^{\theta} \times
		\mathrm{N}\qty(
		Y_{ij} \Big|\mathbf{X}_{ij}\beta_{y,l \mid G_i}
		+\psi_{l \mid G_i}\gamma^{Z_i}_{G_i},\sigma_{l \mid G_i}^2
		),
	\end{align*}
	and then sample $H_{ij}$ from this $l=1,\dots,L$ categorical distribution.

	\subsection{Update of $w^{\phi}_k$ and $v^{\phi}_k$}
	
	Let $v^{\phi}_K=1$. Given $\alpha_{\phi}$ and $G_i$, draw $v^{\phi}_k$ for $k=1,\ldots,K-1$ from
	\begin{align} 
		v^{\phi}_k \sim \text{Be}\left(1 + \sum_{i=1}^{n} \mathbbm{1}(G_i = k), \alpha_{\phi} + \sum_{i=1}^{n} \mathbbm{1}(G_i > k)\right).
	\end{align}
	Then update $w^{\phi}_k = v^{\phi}_k \prod_{j=1}^{k-1} (1 - v^{\phi}_j)$.

	\subsection{Update of $w^{\theta}_{l \mid k}$ and $v^{\theta}_{l \mid k}$}
	
	For each class $k$, let $v^{\theta}_{L \mid k}=1$. Given $\alpha_{\theta \mid k}$ and $H_{ij}$, draw  $v^{\theta}_{l \mid k}$ for $l=1,\ldots,L-1$ from
	\begin{align} 
		v^{\theta}_{l \mid k} &\sim \text{Be}\left(1 + \sum_{i=1}^{n}\sum_{j=1}^{N_i} \mathbbm{1}(H_{ij} = l, G_{i} = k), \alpha_{\theta \mid k} + \sum_{i=1}^{n}\sum_{j=1}^{N_i} \mathbbm{1}(H_{ij} > l, G_{i} = k) \right). 
	\end{align}
	Then update $w^{\theta}_{l \mid k} = v^{\theta}_{l \mid k} \prod_{j=1}^{l-1} (1 - v^{\theta}_{j \mid k})$ for $k=1,\ldots,K$.

	\subsection{Update of $\alpha_{\phi}$ and $\alpha_{\theta \mid k}$}
	
	Assuming a common conjugate prior $\alpha_{\phi}, \alpha_{\theta \mid k} \sim \mathrm{Ga}(a_\alpha, b_\alpha)$, update the concentration parameters $\alpha_{\phi} $ and $ \alpha_{\theta \mid k}$ for $k=1,\ldots,K$:
	\begin{align*} 
		\alpha_{\phi} &\sim \mathrm{Ga}\left( a_\alpha + K - 1, b_\alpha - \sum_{k=1}^{K - 1} \ln(1 - v^{\phi}_k)  \right), \\
		\alpha_{\theta \mid k} &\sim \mathrm{Ga}\left( a_\alpha + L - 1, b_\alpha - \sum_{l=1}^{L - 1} \ln(1 - v^{\theta}_{l \mid k}) \right). 
	\end{align*}

	\subsection{Update of $(\beta_{u,k}, \tau_k)$}
	\noindent
	For each upper-level component $k$, let $n_k$ be the number of subjects $i$ with $G_i=k$.  Denote $U_k$ and $\mathbf{X}_k$ the stacked $U_i$ and covariates for those subjects.  We have:
	\begin{align*}
		\tau_k^2 \mid - &\sim
		\mathrm{IG}\qty(
		a_\tau + \tfrac{n_k}{2},
		b_\tau + \tfrac12 \sum_{i : G_i = k}\qty(U_i - \mathbf{X}_i\beta_{u,k} - \gamma^{Z_i}_k)^{2}
		),\\
		\beta_{u,k}\mid - &\sim
		\mathrm{MVN}\qty(
		\Sigma_k\mathbf{X}_k^\top \qty(U_k - \boldsymbol{\gamma}_{n_k}),
		\tau_k^2 \Sigma_k
		),
		\quad
		\Sigma_k = \qty(\mathbf{X}_k^\top \mathbf{X}_k + \tau_k^2 \Sigma_{\beta_u}^{-1})^{-1}, 
	\end{align*}
	where $\boldsymbol{\gamma}_{n_k}=(\gamma_k^{Z_{k_1}},\ldots,\gamma_k^{Z_{k_{n_k}}})^\top $ is a column vector of size $n_k$ and $k_1,\ldots,k_{n_k}$ are index of subjects with $G_i=k$.
	
	\subsection{Update of $(\beta_{y,l \mid k}, \sigma_{l \mid k})$}
	\noindent
	Similarly, for the recurrent part, each pair $(l \mid k)$ uses data from events where $G_i=k$ and $H_{ij}=l$.  Let $n_{l \mid k}$ be that count, $Y_{l \mid k}$ the stacked responses, and $\mathbf{X}_{l \mid k}$ the corresponding covariates.  Then
	\begin{align*}
		\sigma_{l \mid k}^2 \mid - &\sim
		\mathrm{IG}\qty(
		a_\sigma + \tfrac{n_{l \mid k}}{2},
		b_\sigma + \tfrac12 \sum_{i,j: G_i = k,H_{ij}=l}\qty(Y_{ij} 
		- \mathbf{X}_{ij}\beta_{y,l \mid k} - \psi_{l \mid k}\gamma^{Z_i}_k)^{2}
		), \\
		\beta_{y,l \mid k}\mid - &\sim
		\mathrm{MVN}\qty(
		\Sigma_{l \mid k}\mathbf{X}_{l \mid k}^\top \qty(Y_{l \mid k} - \psi_{l \mid k}\boldsymbol{\gamma}_{n_{l \mid k}}),
		\sigma_{l \mid k}^2 \Sigma_{l \mid k}
		),
		\quad
		\Sigma_{l \mid k} = \qty(\mathbf{X}_{l \mid k}^\top \mathbf{X}_{l \mid k} 
		+ \sigma_{l \mid k}^2 \Sigma_{\beta_y}^{-1})^{-1}.
	\end{align*}
	where $\boldsymbol{\gamma}_{n_{l \mid k}}=(\gamma_k^{Z_{k_1}},\ldots,\gamma_k^{Z_{k_{n_{l \mid k}}}})^\top $ is a column vector of size $n_{l \mid k}$ and $k_1,\ldots,k_{n_{l \mid k}}$ are index of subjects with $G_i=k$ and $H_{ij}=l$.
	
	\subsection{Update of $\boldsymbol{\gamma}_k$}
	\noindent
	We update $\gamma_k^0$ and $\gamma_k^1$ one by one. 
	For the frailty parameter, for each $k$:
	\begin{align*}
		\gamma^{z}_k \mid - \sim \mathrm{N}\qty(\mu_k,s_k),
	\end{align*}
	where
	\begin{align*}
		s_k^{-1} &= \frac{1}{(1-\rho^2) \sigma_{\gamma_z}^2}
		+ \frac{n_k^z}{\tau_k^2}
		+ \sum_{l=1}^{L} \frac{n_{l \mid k}^z \psi_{l \mid k}^{2}}{\sigma_{l \mid k}^2}, \\
		\mu_k &= s_k \left(\frac{\mu_{\gamma_z}}{(1-\rho^2)\sigma_{\gamma_z}^2} + \frac{\rho(\gamma^{1-z}_k-\mu_{\gamma_{1-z}})}{(1-\rho^2)\sigma_{\gamma_z}\sigma_{\gamma_{1-z}}}
		+ \frac{1}{\tau_k^2}\sum_{i: Z_i=z, G_i = k}\qty(U_i - \mathbf{X}_i\beta_{u,k}) \right.\\
		& \left. \quad\quad\quad\quad + \sum_{l=1}^{L}\frac{\psi_{l \mid k}}{\sigma_{l \mid k}^2}
		\sum_{i,j: Z_i=z, G_i = k, H_{ij}=l}\qty(Y_{ij} - \mathbf{X}_{ij}\beta_{y,l \mid k}) \right),
	\end{align*}
	where $n_{k}^z$ be the count of units with $Z_i=z$ and $G_i=k$, and $n_{l \mid k}^z$ be the count of units with $Z_i=z$, $G_i=k$ and $H_{ij}=l$.
	
	\subsection{Update of $\psi_{l \mid k}$}
	\noindent
	Finally, the modulation term $\psi_{l \mid k}$ capturing how $\gamma^{Z_i}_k$ contributes the recurrent-event likelihood is updated by:
	\begin{align*}
		\psi_{l \mid k}\mid - \sim \mathrm{N}\qty(\mu_{l \mid k},s_{l \mid k}),  
	\end{align*}
	where
	\begin{align*}
		s_{l \mid k}^{-1} = \frac{1}{\sigma_\psi^2}
		+ \frac{\sum_{i,j\in G_i=k,H_{ij}=l}(\gamma^{Z_i}_k)^2}{\sigma_{l \mid k}^2},  ~~~
		\mu_{l \mid k} = s_{l \mid k}\qty(\frac{\mu_\psi}{\sigma_\psi^2}
		+ \sum_{i,j\in G_i=k,H_{ij}=l}
		\frac{\gamma^{Z_i}_k \qty(Y_{ij} - \mathbf{X}_{ij}\beta_{y,l \mid k})}{\sigma_{l \mid k}^2} ).
	\end{align*}
	
	\subsection{Compute estimands}
	Given all model parameters, we compute $\kappa_{t, r}(z,x,\boldsymbol{\gamma})$
	and $\eta_{r}(z,x,\boldsymbol{\gamma})$. They are not easy to compute in closed form due to the nonparametric nature of the proposed prior. Therefore we approximate them by the Monte Carlo simulation within each MCMC iteration. For example, we approximate $\kappa_{t, r}(z,x,\boldsymbol{\gamma}) \approx 1/B\sum_{b=1}^{B}N_i^{(b)}(t)$ with $B$ simulated samples of $N_i^{(b)}(t)$ from its posterior predictive distribution.

	\section{Contour plots for SANR}
	\label{sec:sensitivity_analysis_supp}
\begin{figure*}
		\centering
		\begin{subfigure}[b]{14cm}
			\centering
			\includegraphics[width=14cm]{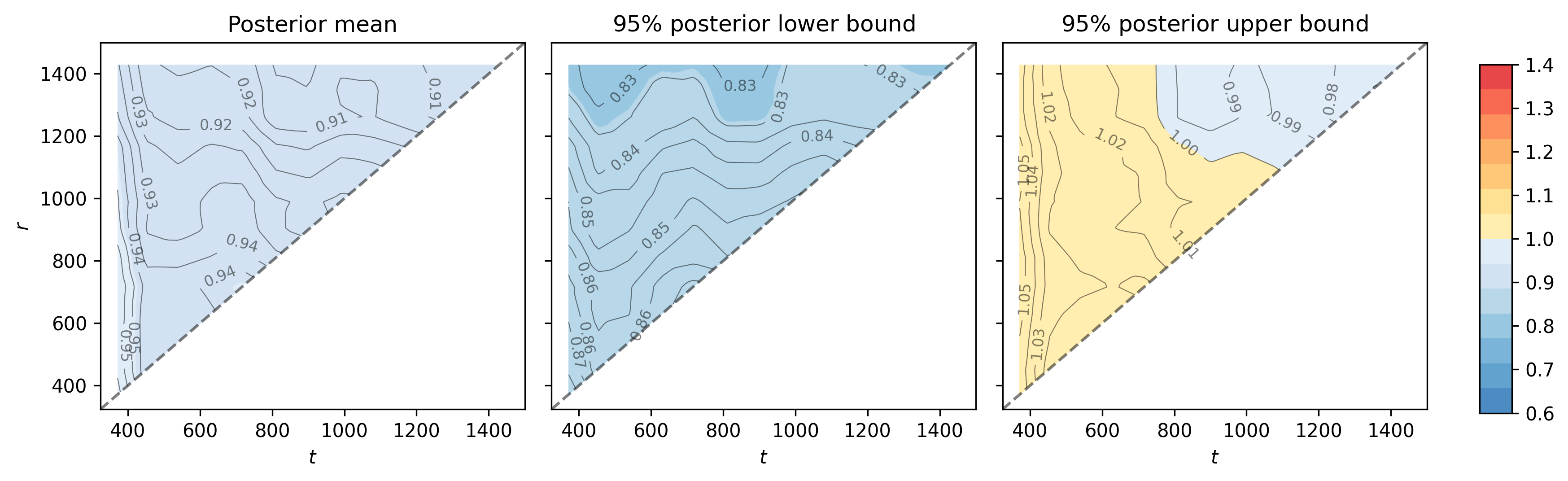}
		\end{subfigure}

		\begin{subfigure}[b]{14cm}
			\centering
			\includegraphics[width=14cm]{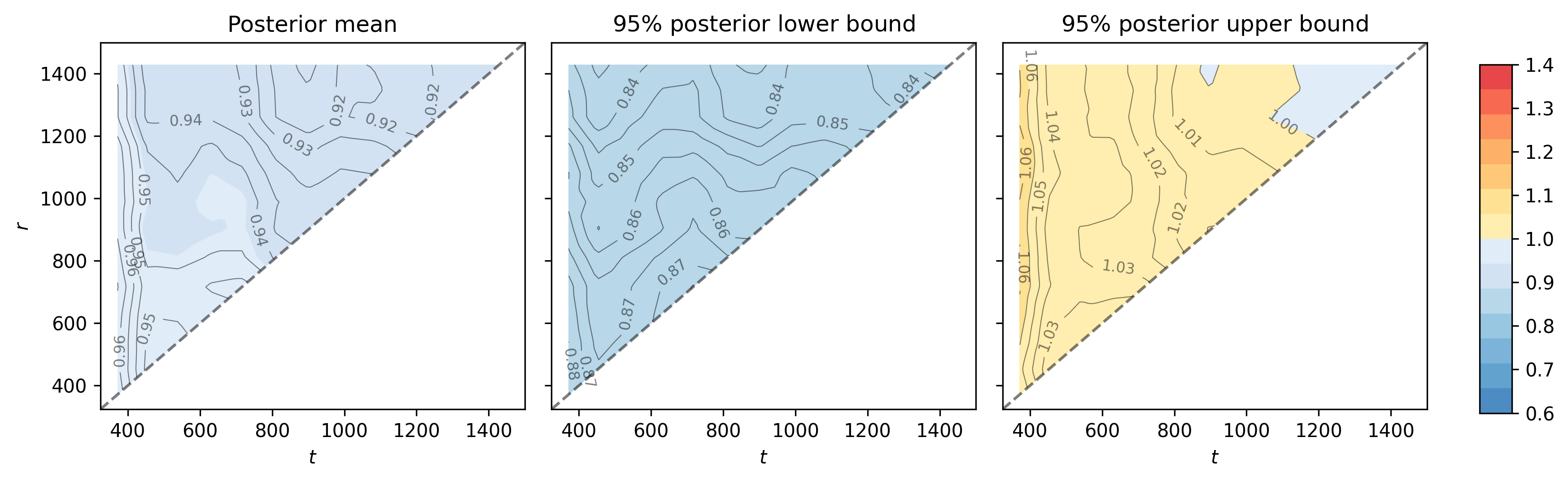}
		\end{subfigure}

		\begin{subfigure}[b]{14cm}
			\centering
			\includegraphics[width=14cm]{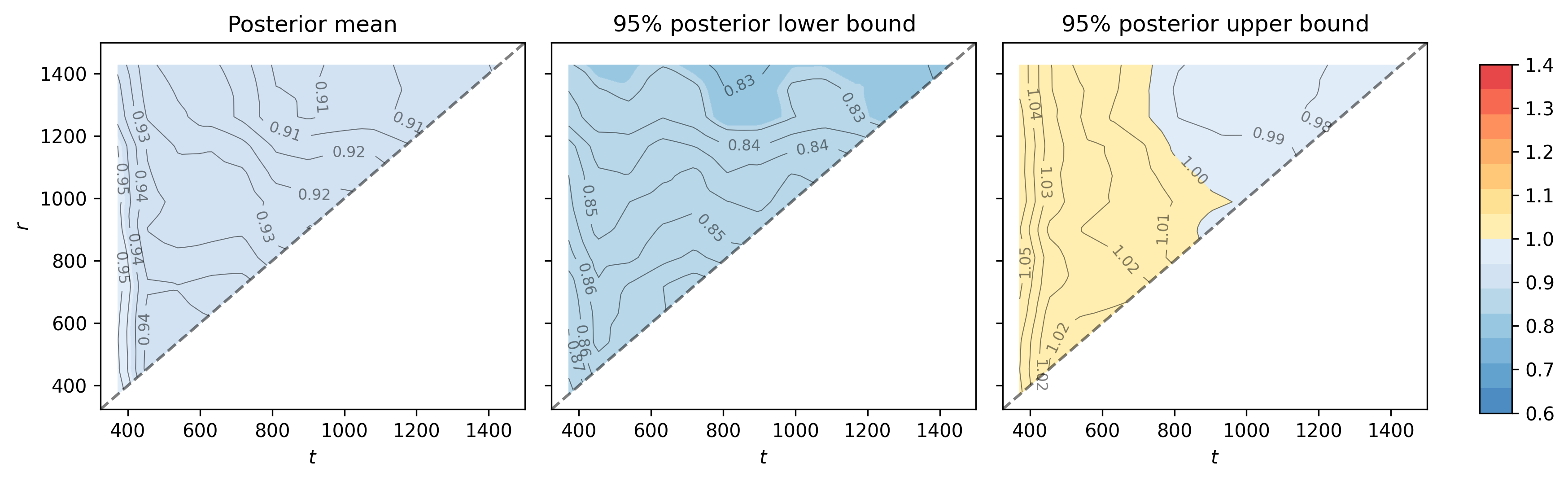}
		\end{subfigure}

		\begin{subfigure}[b]{14cm}
			\centering
			\includegraphics[width=14cm]{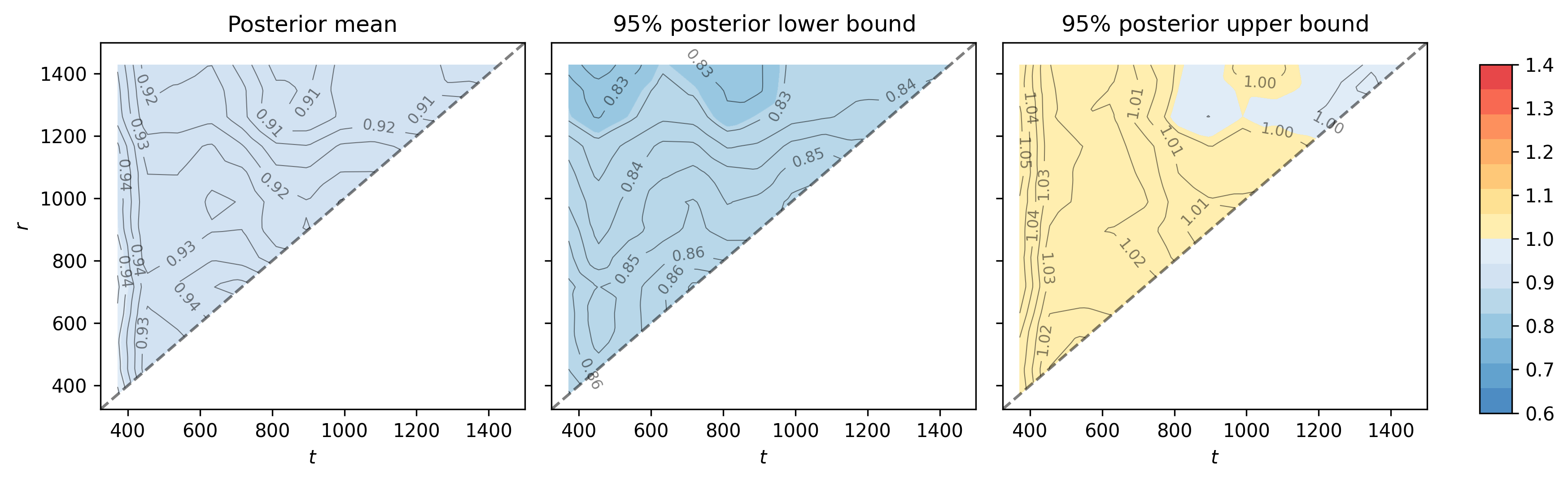}
		\end{subfigure}
		
		\begin{subfigure}[b]{14cm}
			\centering
			\includegraphics[width=14cm]{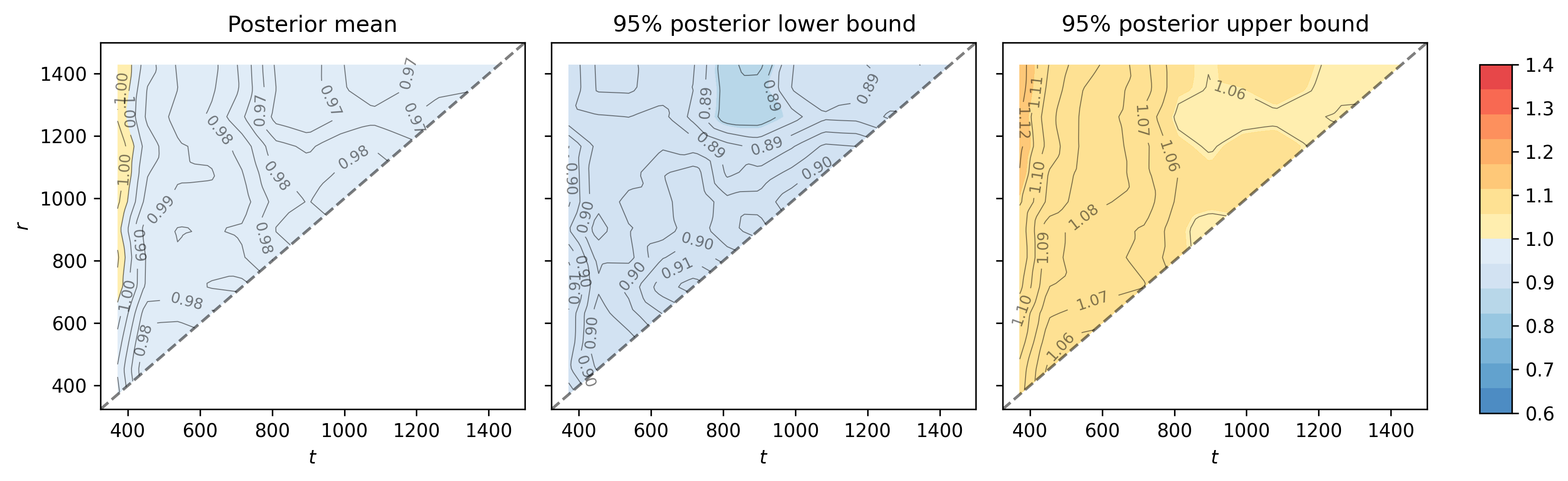}
		\end{subfigure}

		\caption{Contour plots of posterior mean (left), $95\%$ posterior lower bound (middle), and upper bound (right) for SANR. The posterior values are computed for different cut-off values of $t$ and $r$ from $360$ days to $1440$ days with an increment of $90$ such that $t \leq r$, and interpolated between the grid values with a linear spline. The regions with the estimate greater than $1$ are represented using warm colors (yellow to red), while regions with the estimate less than $1$ are indicated using cool colors (blue).}
		\label{fig:sensitivity_analysis_SANR}
	\end{figure*}

	Figure \ref{fig:sensitivity_analysis_SANR} presents contour plots for the SANR across various correlation parameters $\rho$. These plots illustrate how SANR values evolve over a grid of time points $r$ and $t$ ($r\leq t$), consistent with the sensitivity analysis patterns observed in the main manuscript.

\end{document}